\documentclass[11pt]{article}
\usepackage[margin=1in]{geometry}
\usepackage{authblk}
\usepackage[utf8]{inputenc}   
\usepackage[T1]{fontenc}      
\usepackage{hyperref}         
\usepackage{url}              
\usepackage{booktabs}         
\usepackage{amsfonts}         
\usepackage{nicefrac}         
\usepackage{microtype}        
\usepackage{xcolor}           

\usepackage{graphicx}         
\usepackage{subcaption}       
\usepackage{amsmath,amssymb}  
\usepackage{multirow}         
\usepackage{array}            
\usepackage{siunitx}          
\usepackage{caption}
\usepackage{float}
\usepackage{tabularx}
\usepackage{longtable}
\usepackage{placeins}         
\usepackage{comment}
\usepackage{natbib}
\usepackage{pgffor}
\graphicspath{{hazard_2025-08-/}{hazard_2025-09-/}}
\hypersetup{
    colorlinks=true,
    linkcolor=blue!60!black,
    citecolor=blue!60!black,
    urlcolor=blue!60!black,
}

\title{Continuous Flood Nowcasting in South Asia: A Multi-Sensor
Ensemble Remote Sensing Framework for Flood Extent}

\author[1]{Usman Nazir}
\author[2]{Disha Gomathinayagam}
\author[3]{Muhammad Kamran}
\author[1]{Sara Khalid}

\affil[1]{Planetary Health Informatics (PHI) Lab, University of Oxford 
\newline
\texttt{\{usman.nazir, sara.khalid\}@ndorms.ox.ac.uk}}
\affil[2]{Blavatnik School of Government, University of Oxford
\newline
\texttt{disha.gomathinayagam@bsg.ox.ac.uk}}
\affil[3]{PMIU Secretariat, Irrigation Department, Pakistan}  
\date{}

\begin{document}

\maketitle

\begin{abstract}
Pakistan experienced an unusually severe flood season between June and December 2025, with cascading impacts on population, infrastructure, and agriculture. Existing operational flood products (e.g., UNOSAT) provide valuable episode-level snapshots but rarely deliver spatially and temporally continuous inundation maps at near-real-time latency within the country. We present a multi-sensor, ensemble-based remote-sensing framework for \emph{continuous flood nowcasting} in Pakistan that integrates Sentinel-1 SAR, Harmonized Landsat–Sentinel (HLS L30 and S30), MODIS, and VIIRS observations on a harmonized grid in Google Earth Engine. The framework employs a \emph{tiered nowcasting ensemble} that prioritizes higher-resolution sensors (Sentinel-1 and HLS) and falls back to MODIS and VIIRS when necessary, preserving daily continuity of flood extent at each sensor’s native resolution. Applied to the 2025 monsoon period, the system generates near-real-time, spatially consistent inundation maps across Pakistan. As a nowcasting case study, we track the super-flood of 26 August–7 September 2025 day by day, demonstrating the framework’s ability to capture the evolving flood footprint in near real time and extend beyond the temporal limits of episodic mapping products. Validation against GloFAS discharge anomalies and precipitation datasets (CHIRPS v3.0, MSWEP) shows strong agreement with observed hydrometeorological conditions. By integrating nowcast outputs with exposure layers (WorldPop, ESA WorldCover, Giga-HOTOSM), the framework enables rapid estimation of affected populations, cropland, and critical infrastructure, supporting timely disaster response and resilience planning in South Asia.

\end{abstract}

\section{Introduction}

Floods are among the most destructive climate-related hazards in South Asia,
driven by a combination of monsoonal precipitation, glacial melt, tropical
cyclones, and rapid urbanization in exposed flood plains. Timely and spatially
complete flood extent information is essential for humanitarian response,
risk assessment, and climate adaptation planning. However, operational
products such as UNOSAT flood maps are typically produced on an
\emph{episode} basis for specific events and specific countries, leaving
substantial gaps in both space and time. In practice, responders need a
\emph{nowcasting} capability: updated flood maps with latency measured in
days, not weeks, and continuous coverage between episodes rather than
single post-event snapshots.

We develop and apply a cloud-based, multi-sensor ensemble framework for
\emph{continuous flood nowcasting} across Pakistan during the 2025 flood
season, and combine the resulting inundation maps with population, land
cover, and infrastructure data to quantify exposure. Because flood
seasonality differs across the region---five of the six countries
experience the bulk of their flooding during the southwest monsoon, while
Sri Lanka's 2025 flooding was concentrated in the northeast-monsoon and
cyclone-driven late-season window---we adopt \emph{country-specific
observation windows}: JJASO (June--October) 2025 for Pakistan, India, Nepal,
Bangladesh, and Bhutan, and November--December 2025 for Sri Lanka. The
combined reporting window is therefore June--December 2025. To illustrate
the nowcasting behaviour of the framework, we present an in-depth case
study of the Pakistan super-flood of 26 August--7 September 2025
(Section~\ref{sec:nowcast_case}).

\begin{figure*}[!t]
  \centering
  \captionsetup[subfigure]{font=scriptsize,labelformat=empty,
                           justification=centering,skip=2pt}
  \newcommand{\dwidth}{0.19\linewidth}

  \begin{subfigure}[b]{\dwidth}\includegraphics[width=\linewidth]{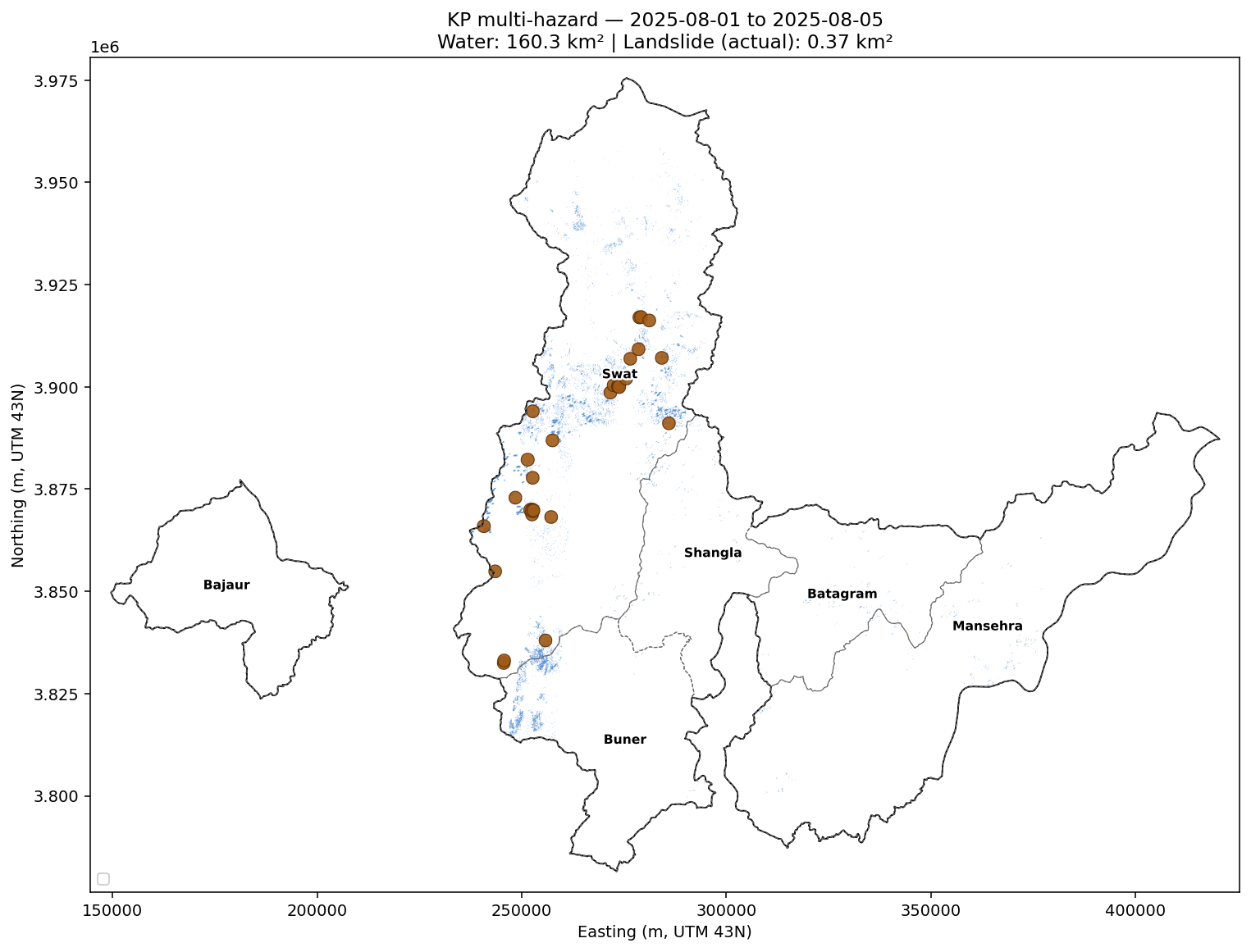}\caption{1--5 Aug}\end{subfigure}\hfill
  \begin{subfigure}[b]{\dwidth}\includegraphics[width=\linewidth]{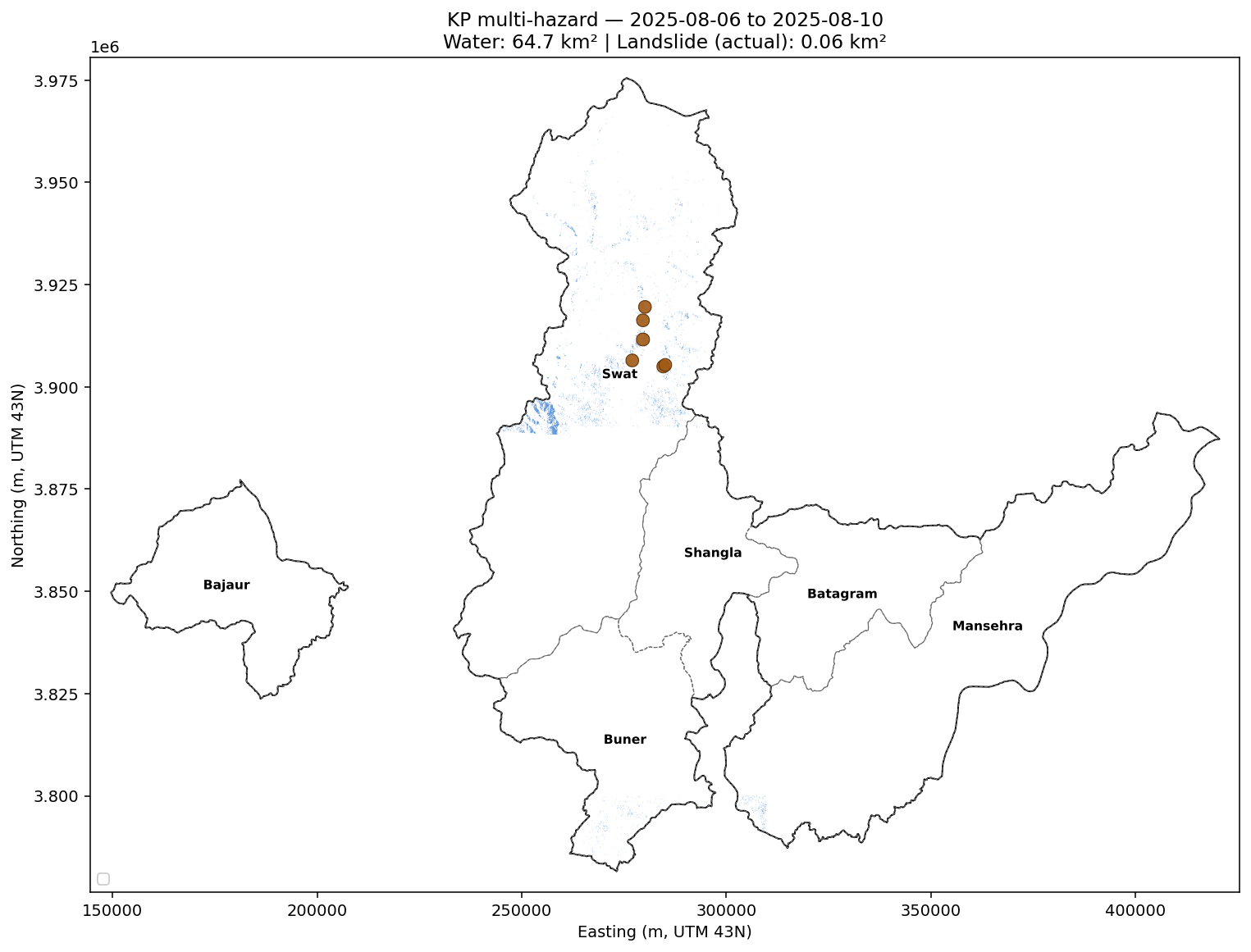}\caption{6--10 Aug}\end{subfigure}\hfill
  \begin{subfigure}[b]{\dwidth}\includegraphics[width=\linewidth]{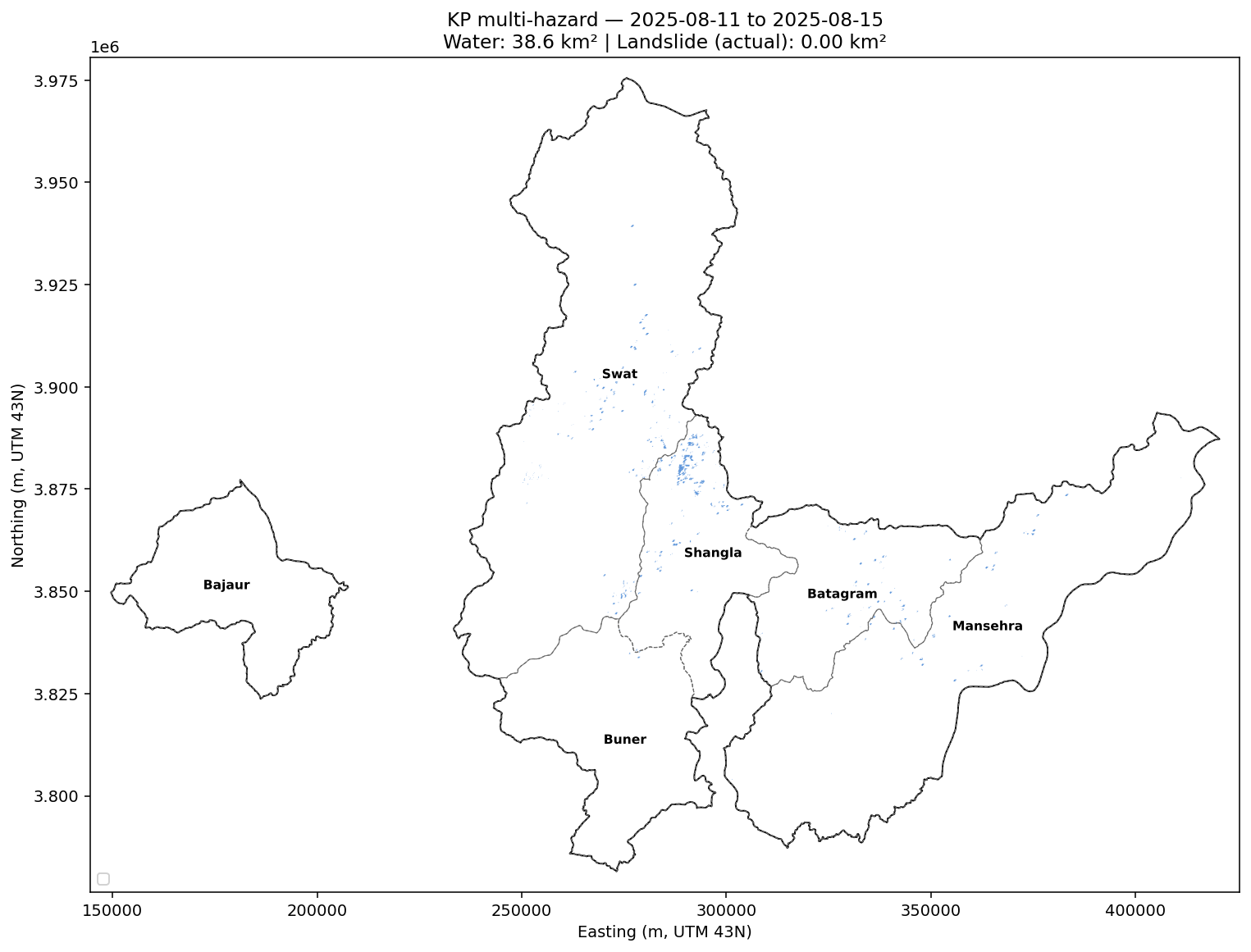}\caption{11--15 Aug}\end{subfigure}\hfill
  \begin{subfigure}[b]{\dwidth}\includegraphics[width=\linewidth]{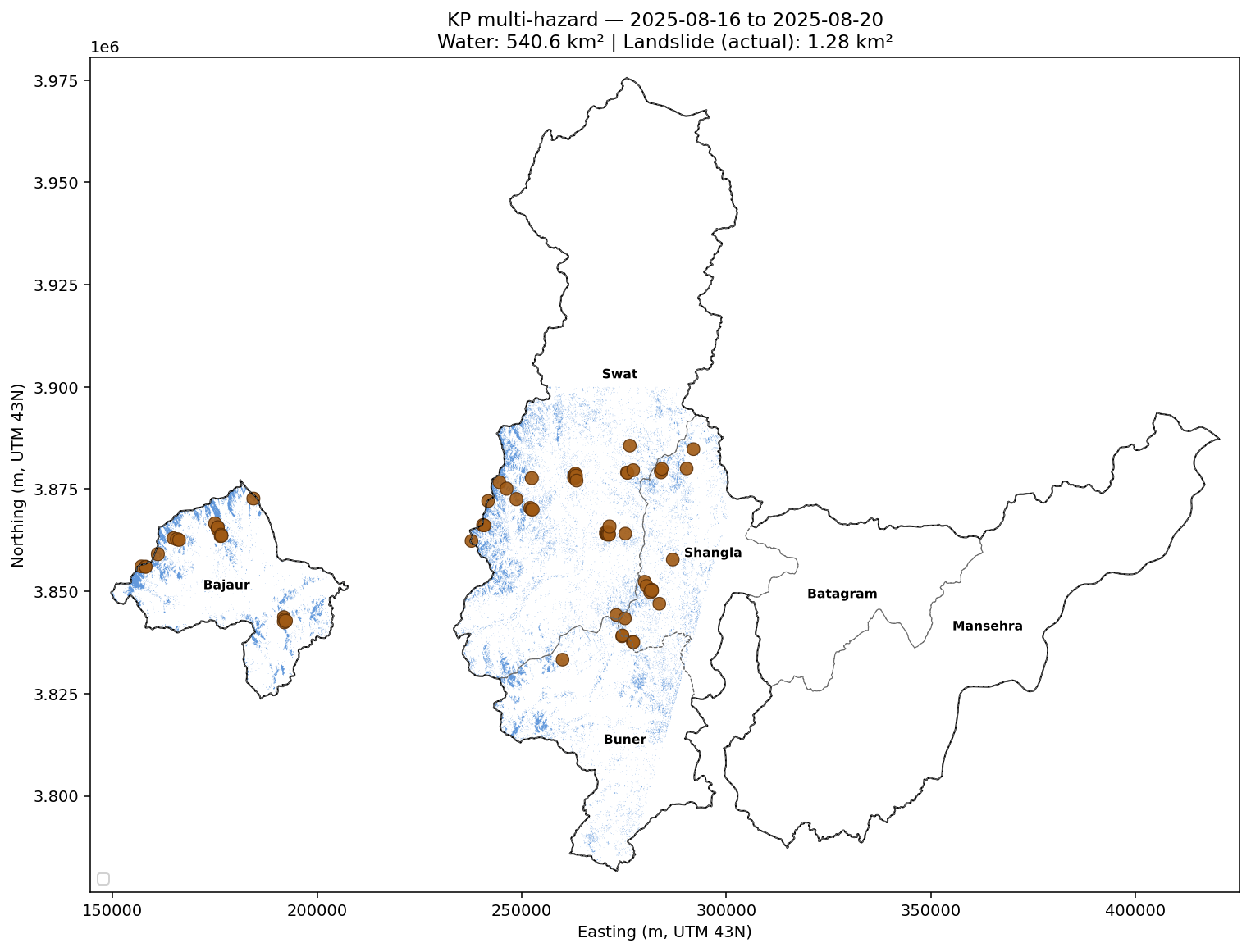}\caption{16--20 Aug \textbf{(cloudburst)}}\end{subfigure}\hfill
  \begin{subfigure}[b]{\dwidth}\includegraphics[width=\linewidth]{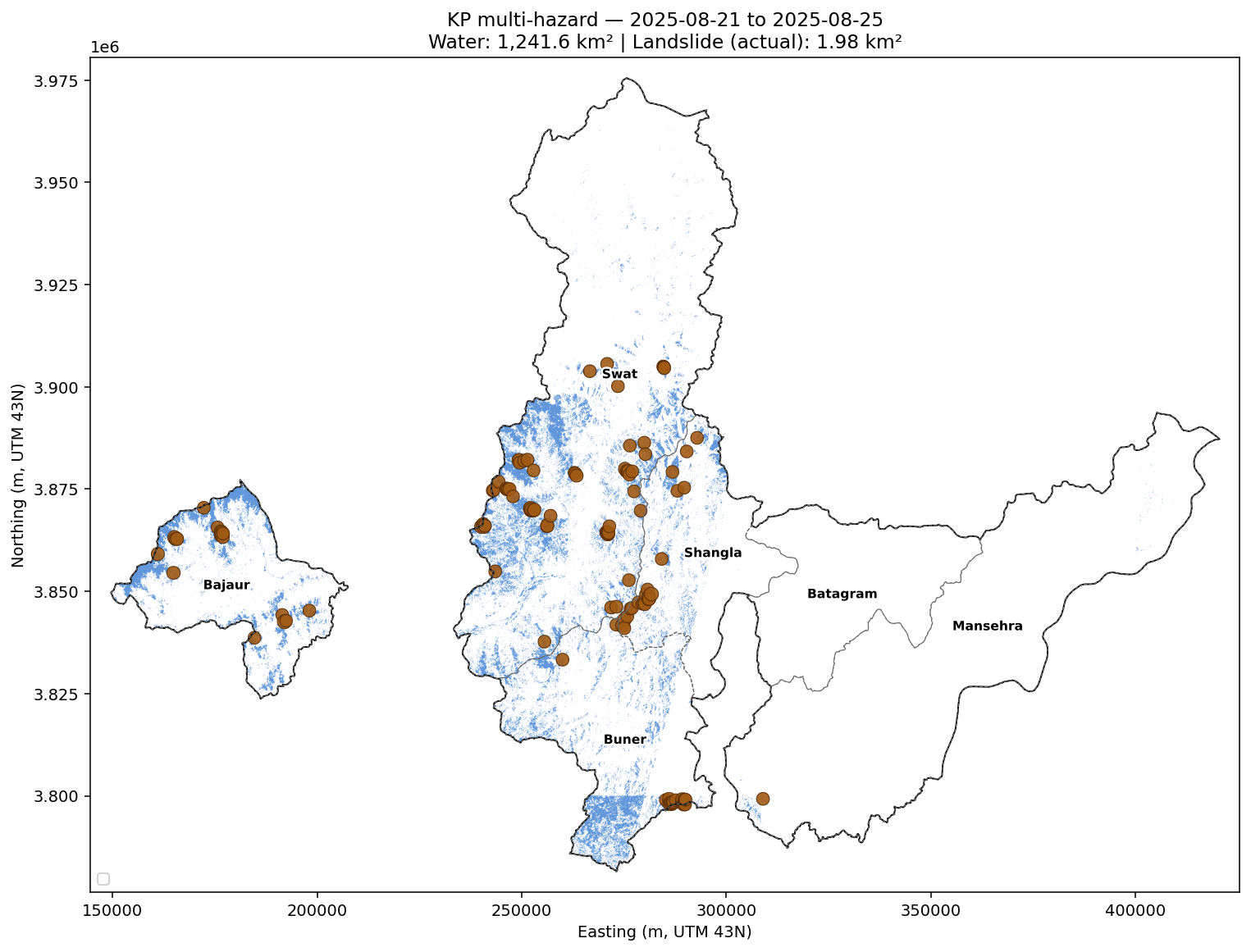}\caption{21--25 Aug \textbf{(peak~1)}}\end{subfigure}

  \vspace{0.5em}

  \begin{subfigure}[b]{\dwidth}\includegraphics[width=\linewidth]{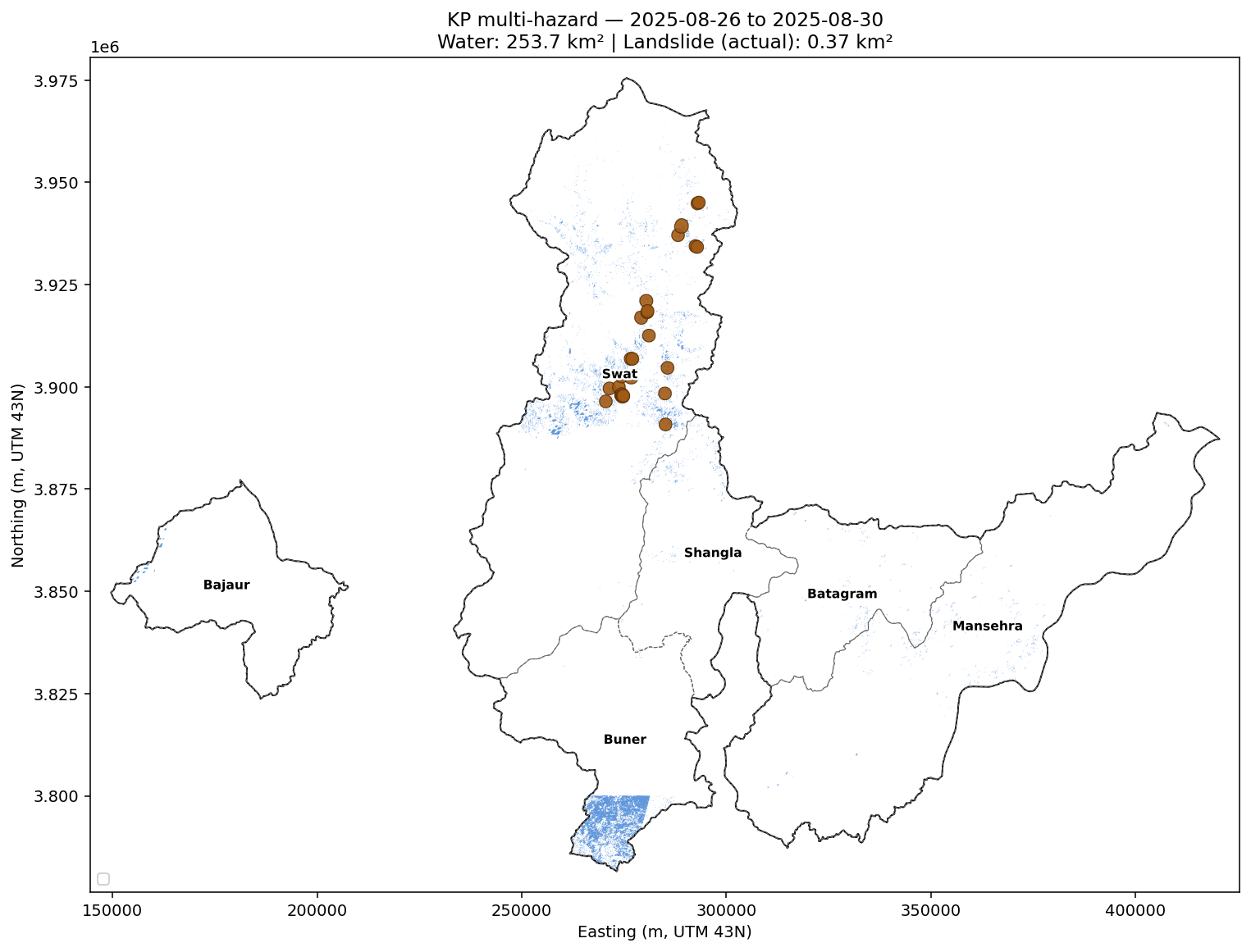}\caption{26--30 Aug}\end{subfigure}\hfill
  \begin{subfigure}[b]{\dwidth}\includegraphics[width=\linewidth]{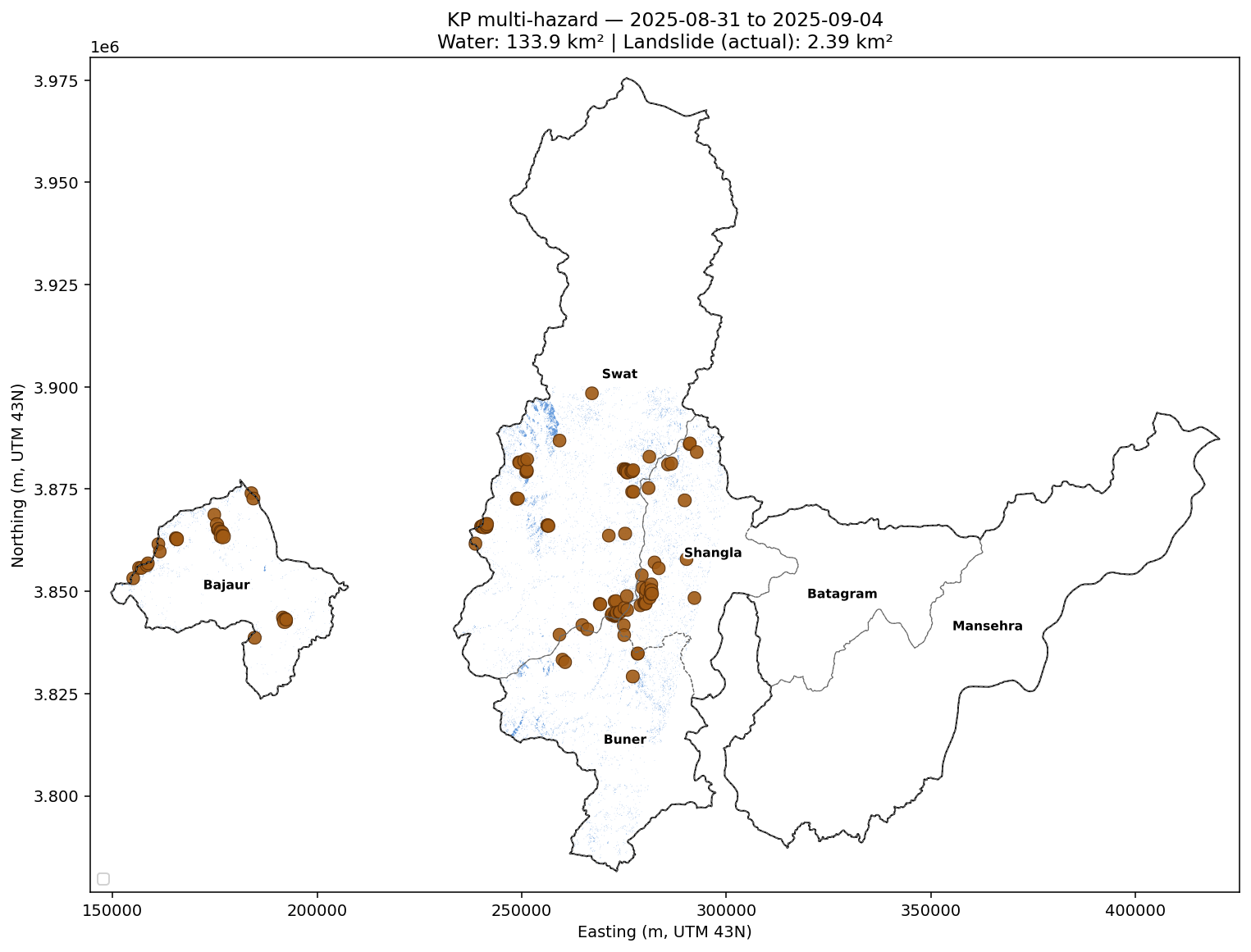}\caption{31 Aug--4 Sep}\end{subfigure}\hfill
  \begin{subfigure}[b]{\dwidth}\includegraphics[width=\linewidth]{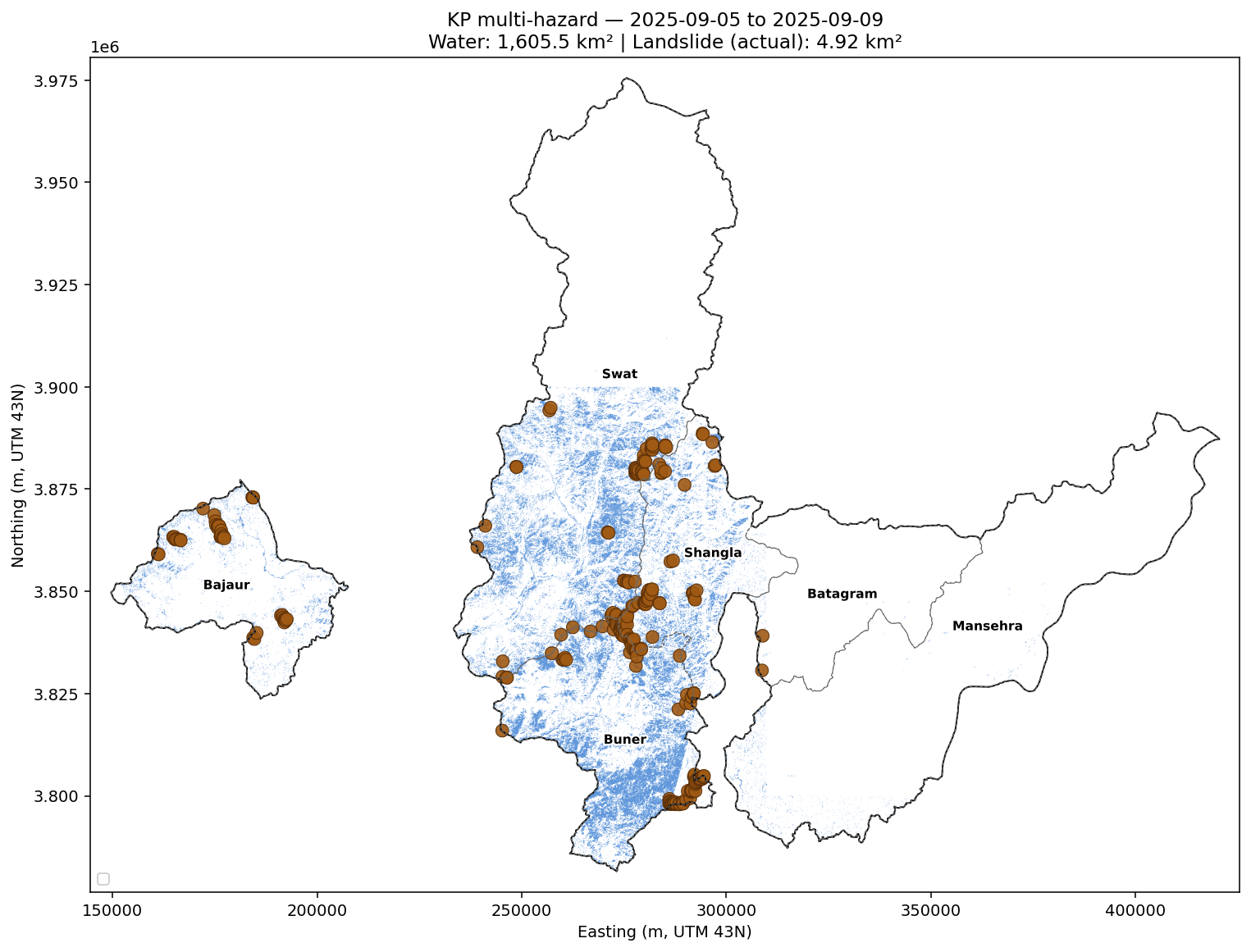}\caption{5--9 Sep \textbf{(peak~2)}}\end{subfigure}\hfill
  \begin{subfigure}[b]{\dwidth}\includegraphics[width=\linewidth]{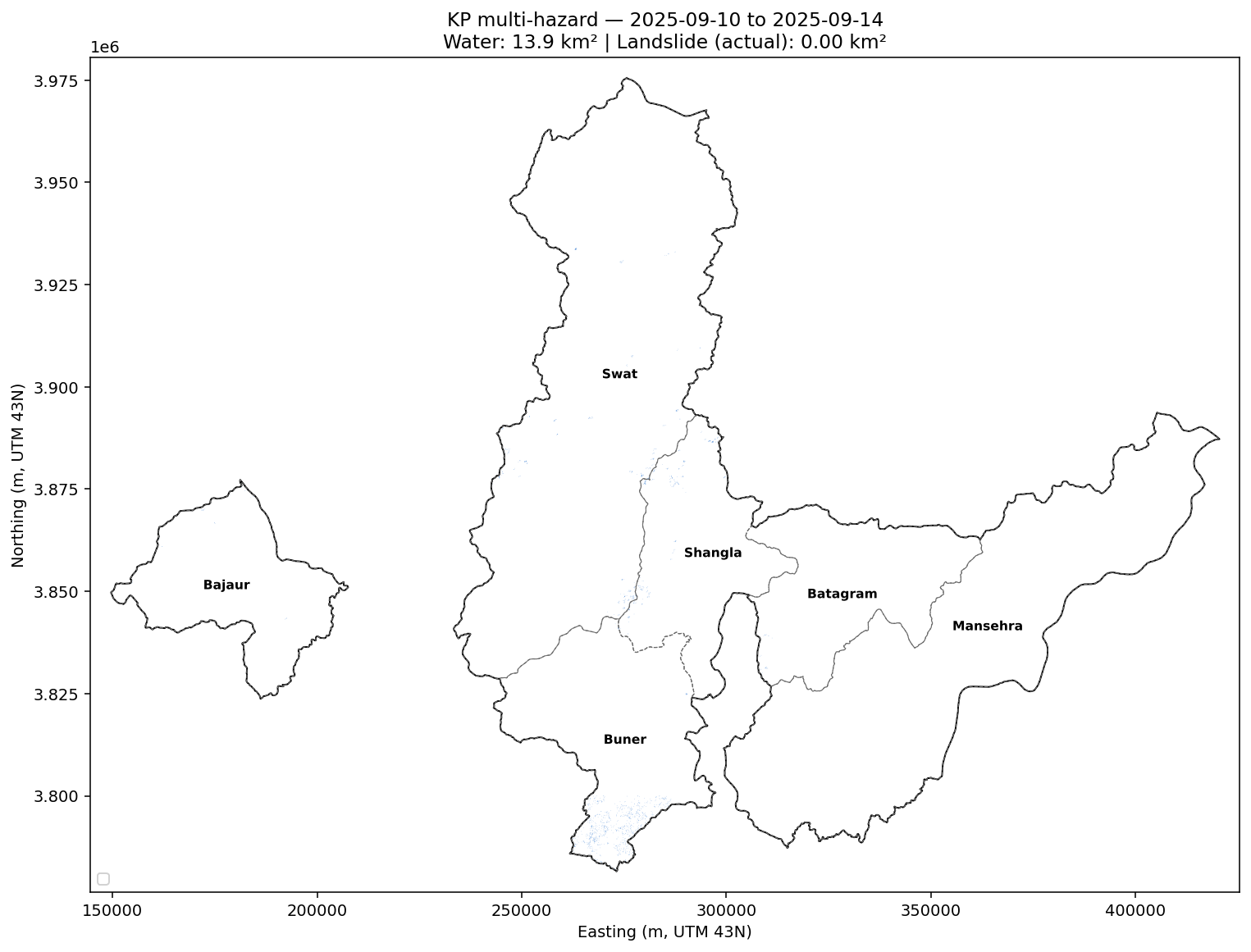}\caption{10--14 Sep}\end{subfigure}\hfill
  \begin{subfigure}[b]{\dwidth}\includegraphics[width=\linewidth]{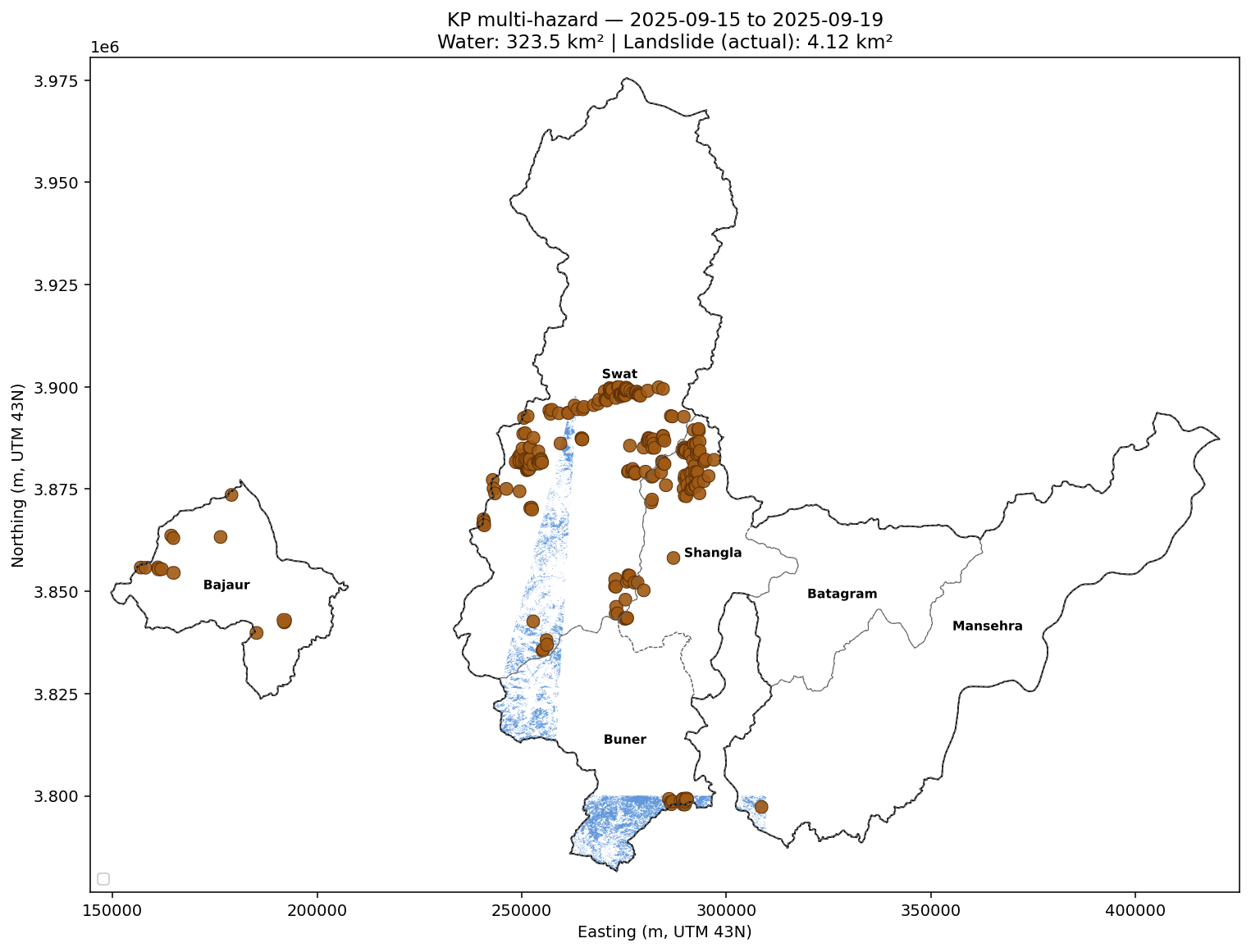}\caption{15--19 Sep}\end{subfigure}

  \vspace{0.7em}

  \begin{subfigure}[b]{0.58\linewidth}
    \includegraphics[width=\linewidth]{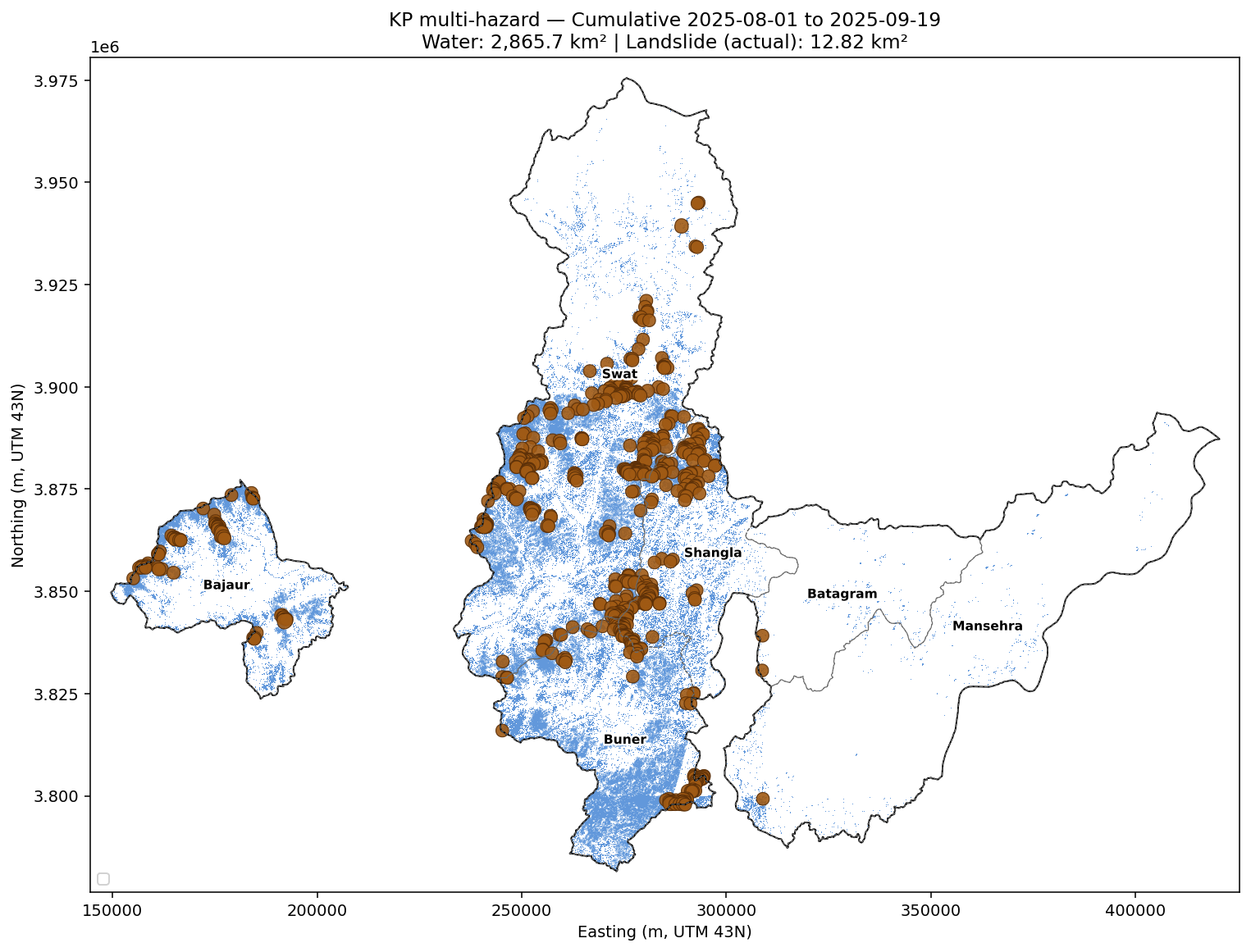}
    \caption{Cumulative 1 Aug -- 19 Sep 2025  \,|\,  Flash/GLOF 2{,}866~km$^2$  \,|\,  landslide 12.8~km$^2$}
  \end{subfigure}

  \caption{%
    Multi-hazard nowcasting in mountainous Khyber-Pakhtunkhwa during the
    August--September~2025 monsoon, demonstrated over six districts:
    Buner, Bajaur, Swat, Shangla, Mansehra and Batagram
    ($\sim$15{,}400~km$^2$ combined).
    Each panel shows the inundation footprint of a 5-day epoch produced
    by our multi-sensor fusion: \textit{rapid-onset flood hazard} (\textcolor[HTML]{1F6DCC}{blue},
    flash floods + glacial-lake-outburst flows) is detected by
    bi-temporal change detection across HLS Sentinel-2, Sentinel-1 SAR,
    HLS Landsat-30, VIIRS and MODIS with $\geq$ 2-of-5 sensor agreement
    (no slope eligibility filter, since mountain water is the target);
    \textit{landslide hazard} (\textcolor[HTML]{A05A14}{brown}) is mapped
    from co-registered HLS L30 + S30 imagery using a NDVI- and
    NBR-loss threshold rule (delta-NDVI~$<-0.15$,
    delta-NBR~$<-0.10$, post-NDVI~$<0.40$) constrained to slopes
    $\geq$~15$^{\circ}$, following Behling~et~al.\ (2014) and
    Mondini~et~al.\ (2011).
    The timeline resolves a clear two-pulse event sequence: the
    16--25~August cloudburst (peak~1, 1{,}241~km$^2$ water in Buner,
    Shangla, Mansehra) followed by a second monsoon surge on
    5--9~September (peak~2, 1{,}604~km$^2$).
    Landslide detections accumulate \emph{after} the floods (peaks at
    epochs~7--8 and~10), capturing the delayed mass-wasting that follows
    saturation of steep slopes.
    The bottom panel combines all 10 epochs (cumulative water
    2{,}866~km$^2$, $\sim$18.6\%~of the region; cumulative landslide
    12.8~km$^2$, $\sim$0.08\%).
    These results demonstrate that the same five-satellite fusion
    framework used for nationwide riverine flooding in Pakistan extends
    naturally to \textbf{flash floods, GLOFs and landslides in
    mountainous terrain}, providing operational nowcasting capability
    across the full spectrum of hydro-meteorological hazards triggered
    by the 2025 monsoon.%
  }
  \label{fig:kp_multihazard}
\end{figure*}

\paragraph{Contributions.}
\begin{itemize}
  \item A reproducible Google Earth Engine pipeline that harmonizes Sentinel-1
    SAR, HLS Landsat-30, HLS Sentinel-2, MODIS, and VIIRS observations and
    operates a two-mode ensemble: (a) a tiered nowcasting ensemble that
    preserves high resolution when Tier-1 sensors are available and uses
    MODIS/VIIRS only as a daily-coverage fallback, and (b) a retrospective
    3-of-5 majority-vote aggregation on a 30\,m grid for seasonal totals
    (Section~\ref{sec:methods}).
  \item A country-specific observation-window strategy (JJASO for five
    countries, November--December for Sri Lanka) that aligns the nowcasting
    cadence with regional flood seasonality.
  \item A nowcasting case study of the Pakistan 26 August--7 September 2025
    event (Section~\ref{sec:nowcast_case}, Figure~\ref{fig:sialkot_daily_flood} and Figure~\ref{fig:kp_multihazard}),
    showing that our framework reproduces the UNOSAT footprint in
    near-real-time and extends it across the full monsoon window.
  \item A 242-event hazard inventory assembled from ReliefWeb, multi-source
    reports, and field expert validation, classifying riverine, flash, urban,
    coastal, cyclone, monsoon, landslide, and compound events
    (Section~\ref{sec:inventory}).
\end{itemize}

\section{Related Work}
\label{sec:related}

\paragraph{Operational flood mapping.}
UNOSAT and the Copernicus Emergency Management Service (EMS) Rapid Mapping
products provide event-triggered flood extents using SAR and optical
imagery, with turnaround in hours to days. They are widely used for
humanitarian response but are activated on a per-event basis, leaving
temporal and spatial gaps across a full monsoon season and unequal country
coverage \citep{unosat,copernicus_ems}.

\paragraph{SAR-based flood detection.}
Sentinel-1 backscatter thresholding is a mature approach for open-water
detection under cloud cover \citep{martinis2015,bauermarschallinger2022}.
Limitations include backscatter ambiguity in urban areas, flooded vegetation,
wet soils, and dry sand. Recent work has explored change-detection and
data-fusion strategies to mitigate these effects.

\paragraph{Optical water indices.}
NDWI \citep{mcfeeters1996}, MNDWI \citep{xu2006}, and AWEI
\citep{feyisa2014} are standard spectral water indicators used with Landsat
and Sentinel-2 observations; Harmonized Landsat--Sentinel (HLS) products
\citep{claverie2018} enable dense time series on a common grid.

\paragraph{Multi-sensor and learned approaches.}
Ensemble fusion of SAR and optical inputs has been shown to reduce
sensor-specific bias. Learned segmentation models trained on benchmarks such
as Sen1Floods11 \citep{bonafilia2020} and the ETCI-2021 challenge
\citep{etci2021} represent a complementary direction; our framework is
deliberately ensemble-and-threshold based to preserve interpretability, run
directly in Google Earth Engine, and avoid labelled-data requirements in
countries where ground truth is sparse.

\paragraph{Hydrological and meteorological reference data.}
We validate against GloFAS discharge reanalysis \citep{harrigan2020} and the
CHIRPS \citep{funk2015} and MSWEP \citep{beck2019} precipitation datasets,
which are standard choices for regional flood and extreme-rainfall analysis.

\section{Proposed Methodology}
\label{sec:methods}

\subsection{Study Design and Temporal Framework}

Flood extent is derived using a multi-sensor remote-sensing framework
integrating synthetic aperture radar (SAR) and optical satellite
observations. A pre-flood (baseline) period and a flood-impact period are
defined using rolling temporal windows. Satellite observations within each
period are filtered and aggregated using median compositing to reduce
noise, residual cloud contamination, and short-term variability. All
analyses are conducted in Google Earth Engine.

\paragraph{Nowcasting windows.}
To support operational nowcasting, the pre-flood baseline and the
flood-impact window are updated on a rolling basis as new Sentinel-1, HLS,
MODIS, and VIIRS acquisitions arrive. Concretely, for a given nowcast
timestamp $t$ we define a pre-flood composite over
$[t-\Delta_{\text{pre}}, t-\Delta_{\text{lead}}]$ and a flood composite
over $[t-\Delta_{\text{flood}}, t]$, with defaults
$\Delta_{\text{pre}}=60$ days, $\Delta_{\text{lead}}=14$ days, and
$\Delta_{\text{flood}}=7$ days. These windows are applied independently
for each sensor and are shifted forward at each update; this produces a
near-real-time flood extent product with latency governed by satellite
revisit, downlink, and Google Earth Engine ingest (on the order of 1--3
days for Sentinel-1 and HLS over our region). Section~\ref{sec:nowcast_case}
illustrates this rolling nowcast behaviour on the Pakistan
26~August--7~September 2025 event.

\paragraph{Study region.}
The study region is defined using FAO GAUL administrative boundaries or
custom subregional geometries. All datasets share a common geographic
coordinate reference system (EPSG:4326). Resampling to a 30\,m common grid
is applied for the retrospective Mode B ensemble
(Section~\ref{sec:ensemble}); for the Mode A tiered nowcast, per-sensor
binary masks are retained at their native resolutions and combined
through the tiered preference rule described in
Section~\ref{sec:ensemble}.

\begin{figure}[H]
  \centering
  \includegraphics[width=\linewidth]{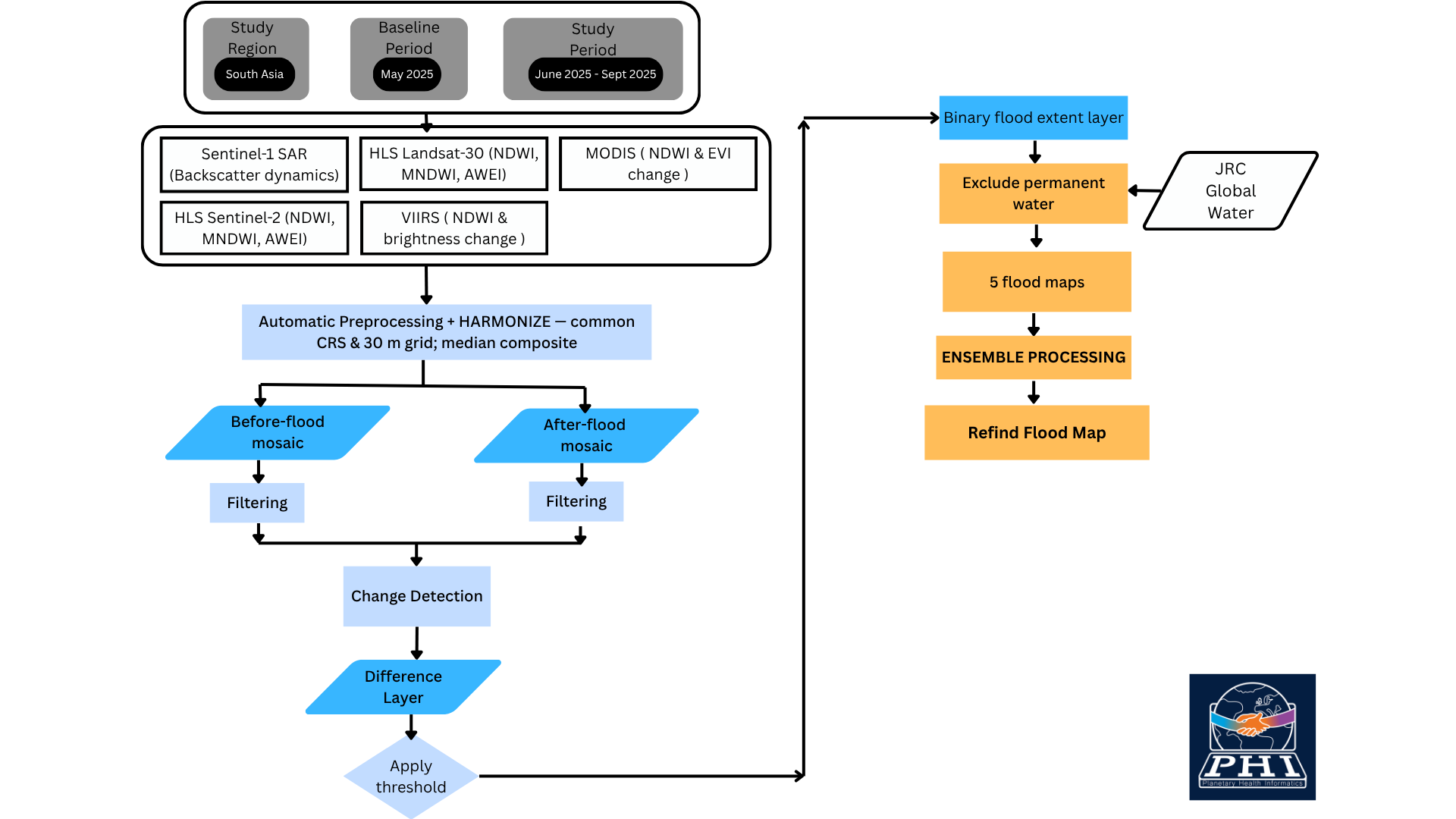}
  \caption{Proposed multi-source, two-mode ensemble methodology for flood
  extent nowcasting and retrospective mapping. The workflow ingests
  multi-sensor satellite data (Sentinel-1 SAR, HLS Landsat-30, HLS
  Sentinel-2, MODIS, and VIIRS) over a defined study region and time
  period, organised into \emph{Tier~1} (S1, L30, S30) and \emph{Tier~2}
  (MODIS, VIIRS). Pre-flood baseline and post-flood mosaics are generated
  for each sensor, differenced, and thresholded to produce per-sensor
  binary flood masks; permanent water bodies are excluded using the JRC
  Global Surface Water dataset. The per-sensor masks then branch into two
  ensemble modes. \textbf{Mode A (Nowcasting, tiered)}: at each nowcast
  timestamp, a pixel is classified using the highest-tier sensors that
  observed it in the current flood window---Tier-1 agreement at 30\,m
  when two or more Tier-1 sensors observed, a single Tier-1 sensor at its
  native resolution when only one observed, MODIS at 500\,m when only
  MODIS observed, and VIIRS at 375--750\,m as a daily-coverage fallback.
  \textbf{Mode B (Retrospective seasonal, 30\,m)}: all per-sensor masks
  are resampled to a harmonised 30\,m grid via nearest-neighbour
  resampling and combined through 3-of-5 majority voting across the full
  observation window. The 30\,m common grid is therefore used for Mode~B
  and for the Tier-1 branches of Mode~A; Mode~A pixels driven by MODIS
  or VIIRS are retained at the driving sensor's native resolution to
  avoid artificial upsampling.}
  \label{fig:methodology}
\end{figure}

\subsection{Multi-Sensor Flood Detection}

Flood detection is performed independently using five satellite systems:
Sentinel-1 SAR (VV/VH backscatter); HLS Landsat-30 (L30); HLS Sentinel-2 (S30);
MODIS Terra/Aqua; and VIIRS. Each dataset is processed separately to produce
binary flood masks.

\paragraph{Sentinel-1 SAR flood detection.}
Sentinel-1 Ground Range Detected (GRD) imagery acquired in Interferometric
Wide (IW) mode is filtered by date and region. Median composites are
generated for both pre-flood and flood periods. Flooded pixels are identified
using a combination of: (i) a backscatter decrease between pre- and
post-flood composites ($\Delta$VV threshold); and (ii) an absolute low
post-flood VV backscatter threshold. Pixels satisfying either condition are
classified as flooded. Permanent water bodies are removed using the JRC
Global Surface Water dataset \citep{pekel2016} (occurrence $\geq 50\%$).

\paragraph{Optical flood detection (HLS Landsat-30 and Sentinel-2).}
HLS imagery is filtered for cloud coverage ($<20\%$) and scaled to surface
reflectance. Median composites are generated for the pre- and post-flood
periods. Three water-related spectral indices are computed: the Normalized
Difference Water Index (NDWI), the Modified NDWI (MNDWI), and the Automated
Water Extraction Index (AWEI). Flood pixels are identified where at least
two of the three indicators exceed predefined change thresholds between
pre- and post-flood composites. Permanent water masking is applied.

\paragraph{MODIS and VIIRS flood detection.}
For MODIS (MOD09A1 and MYD09A1) and VIIRS (VNP09GA), pre- and post-flood
median composites are generated. Flood detection is based on: (i) an
increase in NDWI; and (ii) a decrease in vegetation index (EVI for MODIS)
or surface brightness (VIIRS). Pixels satisfying both spectral conditions
are classified as flooded. Permanent water bodies are excluded.

\subsection{Spatial Harmonization and Two-Mode Ensemble Processing}
\label{sec:ensemble}

Because the five sensors in our framework span native resolutions from
10\,m (Sentinel-1, Sentinel-2) to 750\,m (VIIRS) and have very different
revisit cadences---6--12 days for Sentinel-1, 2--3 days for HLS
(L30$+$S30 combined), and daily for MODIS and VIIRS---a single
aggregation rule cannot serve both near-real-time nowcasting and
full-season retrospective analysis. We therefore operate the ensemble in
\emph{two complementary modes}, both applied after the per-sensor binary
flood masks have been reprojected to a common geographic CRS (EPSG:4326).

\paragraph{Mode A --- Tiered nowcasting ensemble.}
For the nowcasting product, we organize the five sensors into two tiers
reflecting their effective detection accuracy and native spatial
resolution:
\begin{itemize}
  \item \textbf{Tier 1 (primary):} Sentinel-1 SAR (10\,m), HLS
    Landsat-30 (30\,m), HLS Sentinel-2 (30\,m). These are the highest
    spatial resolution and highest-accuracy flood delineators in the
    stack.
  \item \textbf{Tier 2 (daily-coverage fallback):} MODIS
    (250--500\,m), VIIRS (375--750\,m). These provide daily revisit at
    the cost of spatial detail.
\end{itemize}
At each nowcast timestamp $t$, the ensemble is constructed with a tiered
preference rule applied per pixel:
\begin{enumerate}
  \item If two or more Tier-1 sensors observed the pixel within the
    current flood window, a pixel is classified as flooded only when
    \emph{both} observing Tier-1 sensors detect flood. The output is
    written at 30\,m.
  \item If exactly one Tier-1 sensor observed the pixel, that sensor's
    binary detection drives the classification. The output is written at
    the Tier-1 native resolution.
  \item If no Tier-1 sensor observed the pixel but MODIS did, MODIS
    drives the classification and the output is written at 500\,m
    (MODIS native).
  \item If only VIIRS observed the pixel, VIIRS drives the
    classification and the output is written at VIIRS native resolution
    (375--750\,m). VIIRS therefore serves specifically as a daily
    continuity fallback to prevent gaps in the nowcast.
  \item If no sensor observed the pixel within the flood window, the
    pixel is labelled \emph{no observation}.
\end{enumerate}
The nowcasting mosaic is assembled at the native resolution of the
\emph{driving sensor}---we do not artificially upsample MODIS or VIIRS to
30\,m, which would give a false impression of spatial detail. The output
is binary (flooded / not flooded), with a separate observation mask
recording which tier produced each pixel so that downstream exposure
calculations can weight by observation quality if desired.

\paragraph{Mode B --- Retrospective 3/5 majority-vote aggregation.}
For the full-season totals reported in Section~\ref{sec:results_extent}, we use a retrospective majority-vote ensemble over the country-specific
observation window. All five sensors contribute equally. All binary masks
are resampled to a common 30\,m grid using nearest-neighbor resampling,
and then aggregated pixel-wise across the full window:
\begin{itemize}
  \item \textbf{Union:} flooded if detected by any sensor at any time in
    the window.
  \item \textbf{Intersection:} flooded only if detected by all sensors.
  \item \textbf{Majority voting (3/5 and 4/5):} flooded if at least 3 or
    at least 4 sensors detected flood at any time.
  \item \textbf{Confidence map:} number of sensors (0--5) detecting
    flood.
\end{itemize}
The 3/5 majority-vote mosaic is selected as the primary retrospective
flood-extent product, balancing omission and commission errors while
leveraging cross-sensor agreement over the full window.

\paragraph{Why two modes.}
Mode A optimizes \emph{timeliness} at a given date, producing a
best-available inundation map updated every 48 hours with explicit
sensor-driven resolution. Mode B optimizes \emph{aggregate reliability}
over a full monsoon window, pooling all available observations and
requiring multi-sensor agreement before a pixel is counted toward the
seasonal total. 

\paragraph{Permanent water masking.}
In both modes, pixels with JRC Global Surface Water \citep{pekel2016}
occurrence $\geq 50\%$ are excluded from the final flood extent to avoid
mislabelling permanent water bodies as flooded. Flooded area is
calculated using per-pixel area aggregation at each pixel's native
resolution.

\subsection{Methodological Rationale}

By integrating SAR-based backscatter dynamics, multi-index optical water
detection, and coarse-resolution daily revisit from MODIS and VIIRS in a
two-mode ensemble---tiered and resolution-preserving for nowcasting
(Mode A), and 3-of-5 majority-voting on a 30\,m grid for retrospective
aggregation (Mode B)---the framework provides: robust detection under
cloud cover, reduced sensor-specific bias, spatially detailed inundation
mapping when Tier-1 sensors are available, uninterrupted daily continuity
when they are not, and hydrologically consistent flood representation.

\subsection{Software}

To process these diverse datasets efficiently, we utilize Google Earth Engine
\citep{gorelick2017}, a cloud-based platform that enables large-scale
analysis of remote sensing data. Its cloud infrastructure provides enhanced computational speed by
outsourcing processing to Google's servers, thereby eliminating the need to
download and manage raw imagery locally. Results are visualized on an
interactive map and can be exported as georeferenced files for downstream
use. Resulting maps are combined with population and infrastructure data
using QGIS v3.44.

\FloatBarrier

\section{Multi-Source Event Inventory and Classification}
\label{sec:inventory}

Between June and December 2025, a total of 23 flood-related hazard events were documented across Pakistan. These comprised 9 flash flooding events and 9 urban flooding events, representing the dominant hydrometeorological drivers during the study period. In addition, 3 riverine flooding events were recorded, primarily along the Indus floodplain and its tributaries in Sindh and Punjab. A single monsoon flooding event was identified in Badin District, Sindh, and one glacial lake outburst flood (GLOF) was recorded in the Hunza Valley, Gilgit-Baltistan. Geographically, events clustered across Khyber Pakhtunkhwa (flash flooding in Buner, Upper Dir, Swat, Shangla, and Bajaur), Punjab (urban and riverine flooding in Sialkot, Gujrat, Narowal, Sheikhupura, Hafizabad, Mandi Bahauddin, Rajanpur, and Muzaffargarh), and Sindh (riverine, urban, and monsoon flooding in Sukkur, Karachi, and Badin). Event identification varied by source: ReliefWeb documented 6 events, while the wider multi-source compilation — drawing on UNOCHA, Shelter Cluster, national media (Dunya News, Arab News, Express Tribune), and Wikipedia — captured all 23 events. Field-expert validation confirmed all recorded events in Pakistan and led to the addition of 8 urban flooding cases across Punjab, strengthening the completeness and reliability of the final inventory.

\section{Results: Continuous Flood Extents in Pakistan (June--December 2025)}
\label{sec:results_extent}

The total flood extent detected across Pakistan was (114{,}513.0\,km$^{2}$), which was the second largest inundated area observed across the study region over the full June--December 2025 reporting window. 

\FloatBarrier

\begin{figure}[h]
  \centering
  \includegraphics[width=\linewidth]{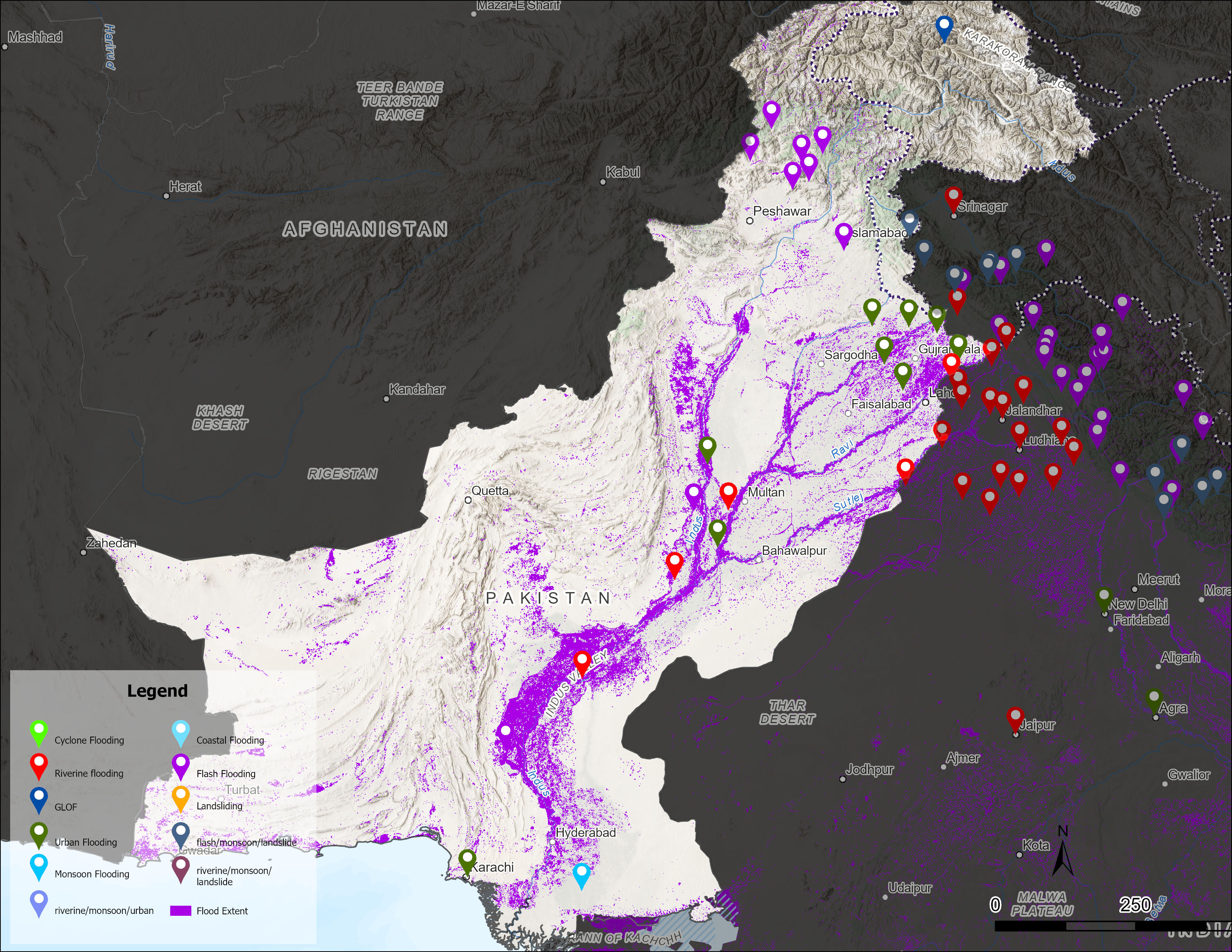}
  \caption{Spatial distribution and impacts of flood hazards across Pakistan. The panel shows the geographic distribution of recorded flood
  events, categorized by flood type (e.g., riverine, urban, flash, coastal, and landslide-related).}
  \label{fig:spatial_hazard}
\end{figure}

\subsection{Nowcasting Case Study 1: Pakistan 26~August--02~September 2025}
\label{sec:nowcast_case}

\begin{figure*}[!t]
  \centering
  \captionsetup[subfigure]{font=scriptsize,labelformat=empty,justification=centering,skip=2pt}
  \newcommand{\dwidth}{0.135\linewidth}

  \begin{subfigure}[b]{\dwidth}\includegraphics[width=\linewidth]{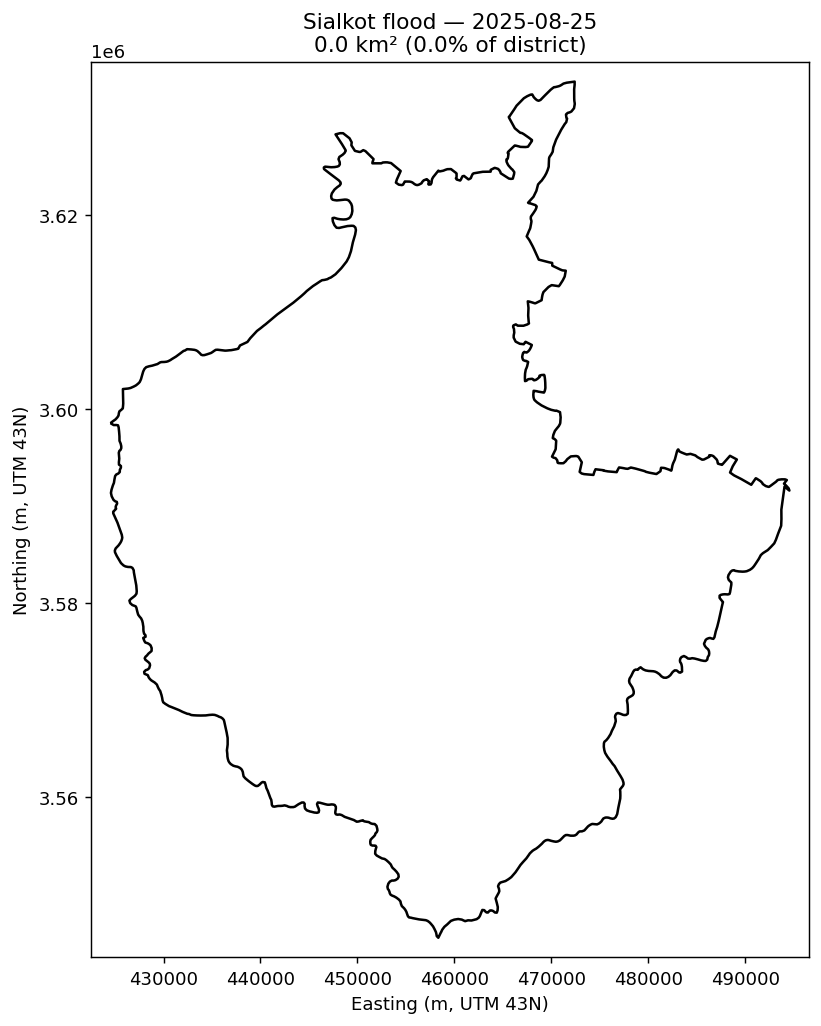}\caption{25 Aug}\end{subfigure}\hfill
  \begin{subfigure}[b]{\dwidth}\includegraphics[width=\linewidth]{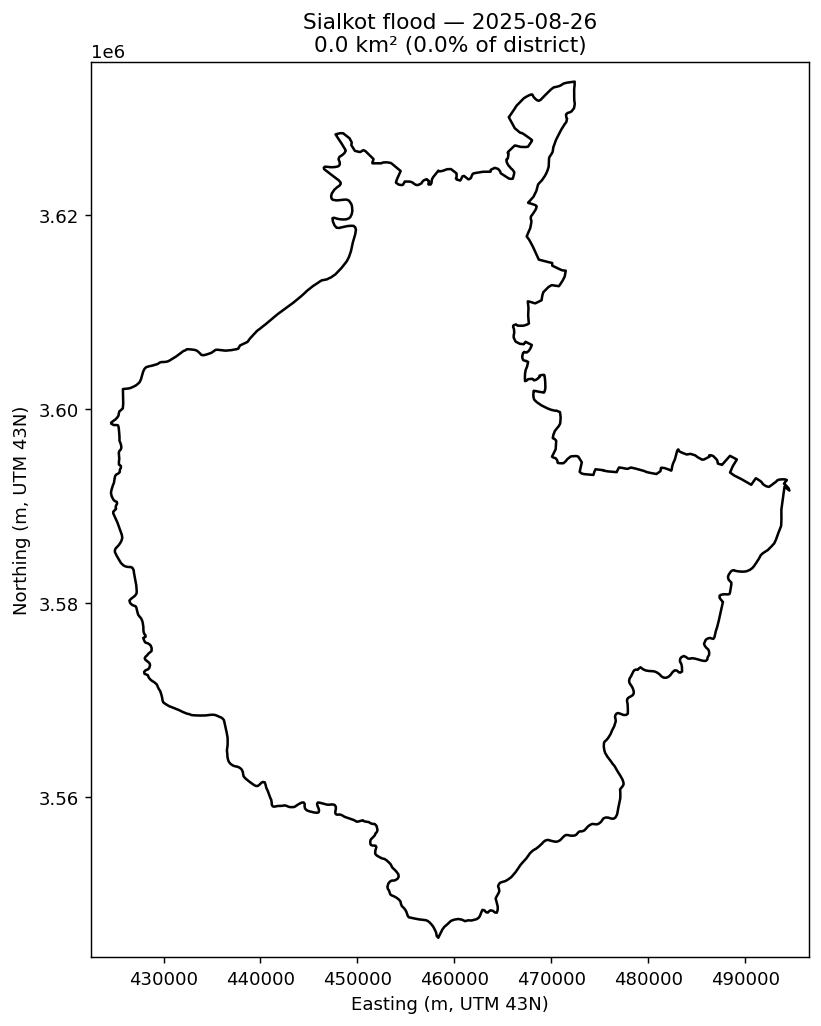}\caption{26 Aug}\end{subfigure}\hfill
  \begin{subfigure}[b]{\dwidth}\includegraphics[width=\linewidth]{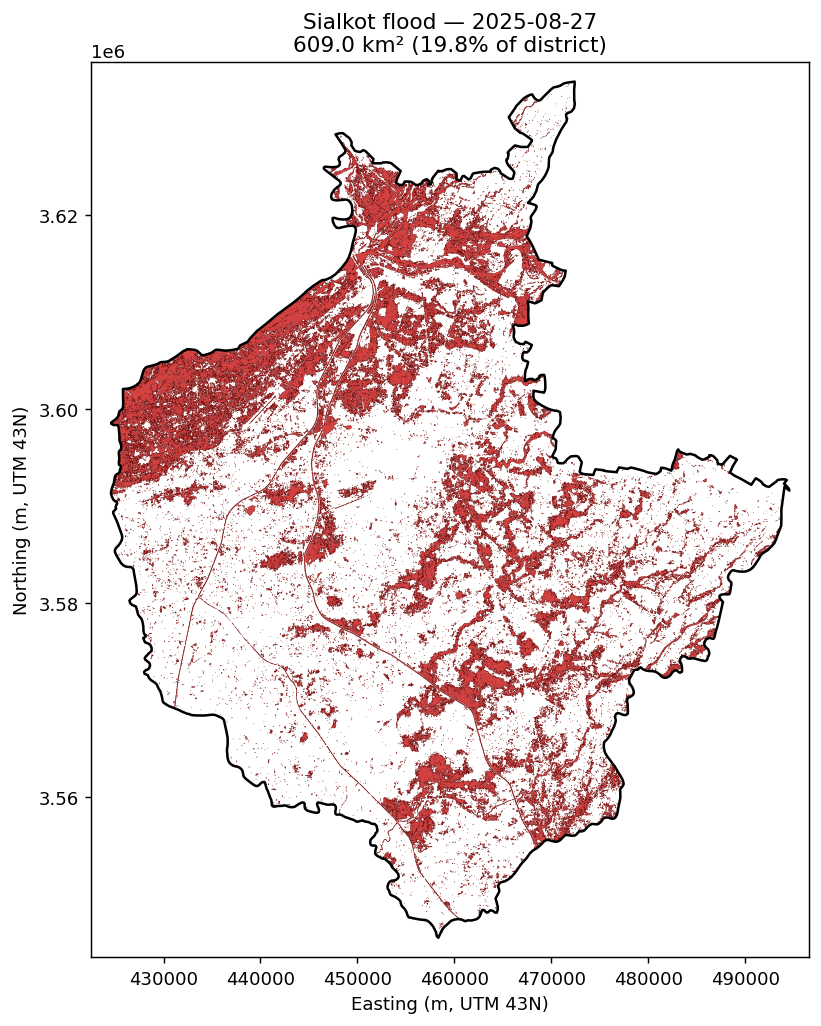}\caption{27 Aug}\end{subfigure}\hfill
  \begin{subfigure}[b]{\dwidth}\includegraphics[width=\linewidth]{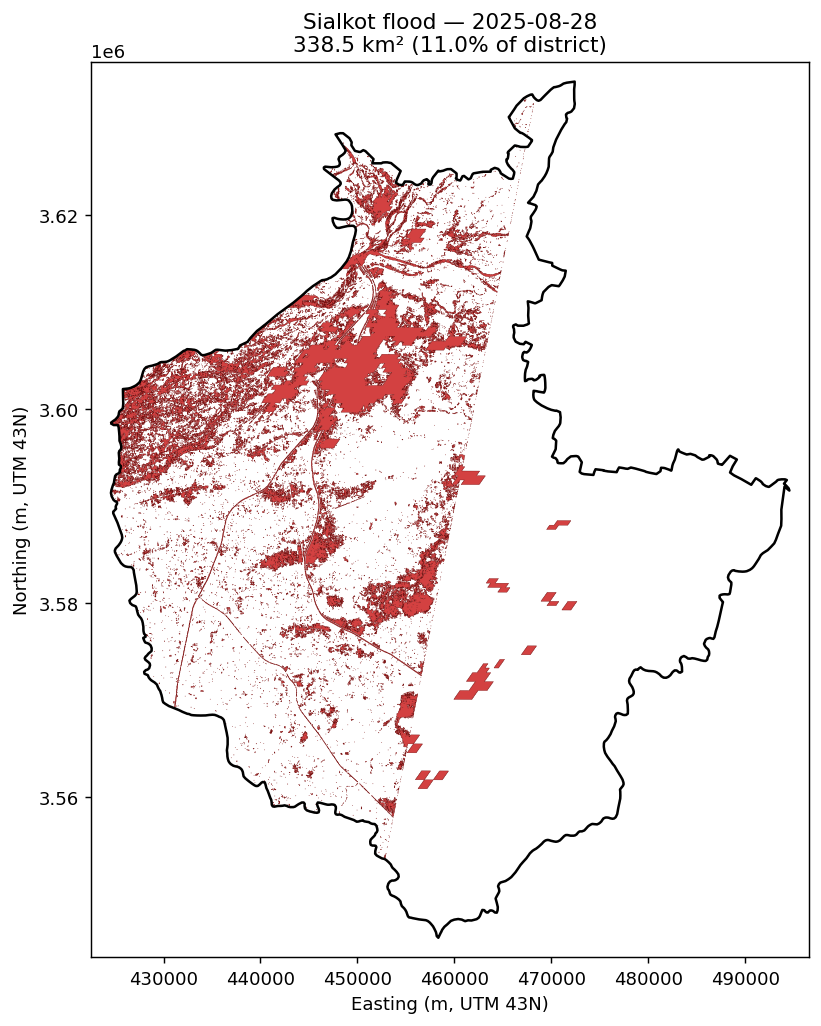}\caption{28 Aug}\end{subfigure}\hfill
  \begin{subfigure}[b]{\dwidth}\includegraphics[width=\linewidth]{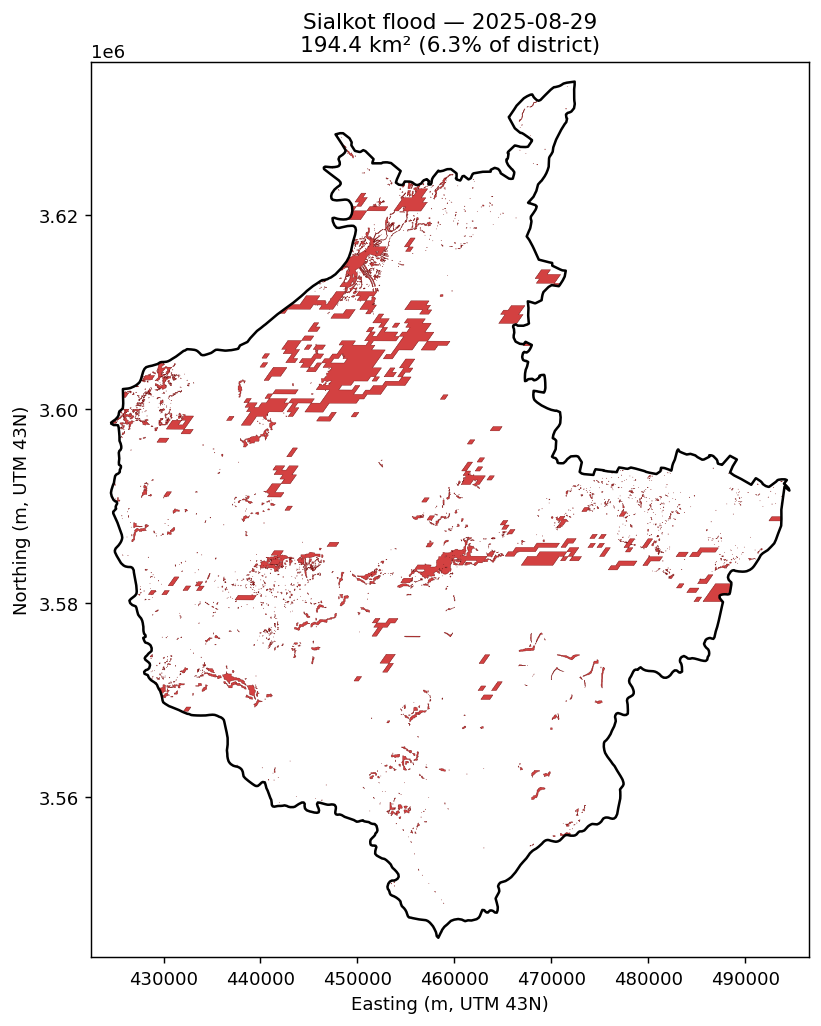}\caption{29 Aug}\end{subfigure}\hfill
  \begin{subfigure}[b]{\dwidth}\includegraphics[width=\linewidth]{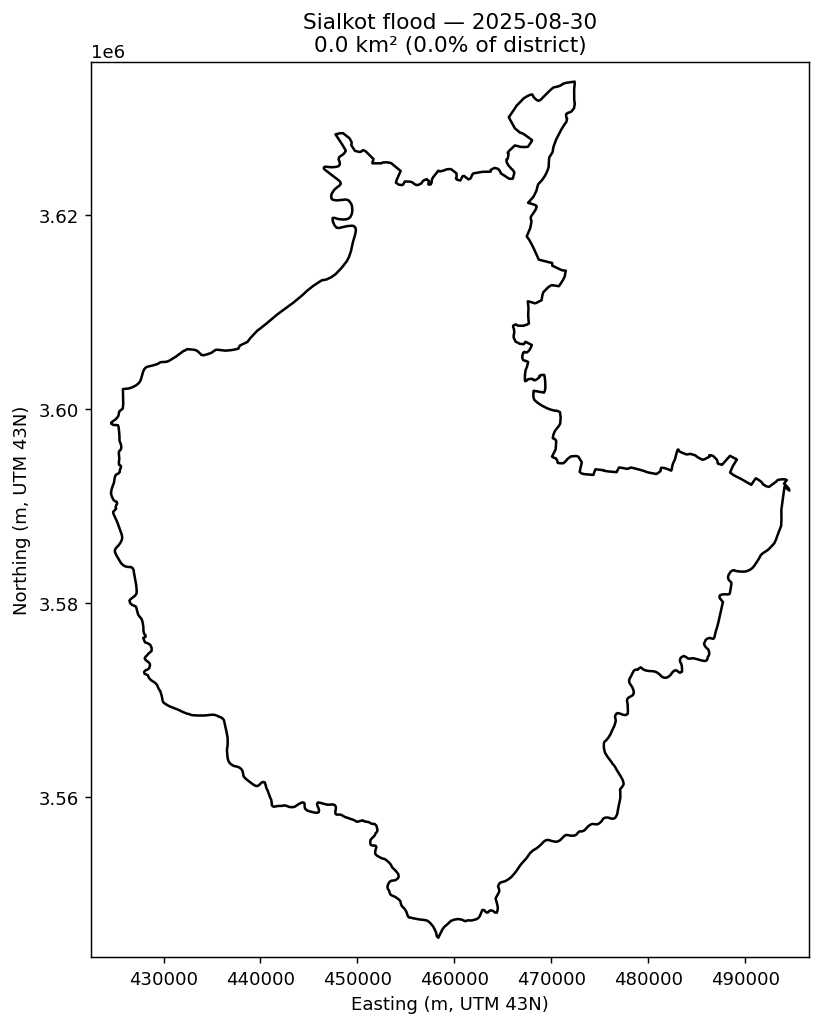}\caption{30 Aug}\end{subfigure}\hfill
  \begin{subfigure}[b]{\dwidth}\includegraphics[width=\linewidth]{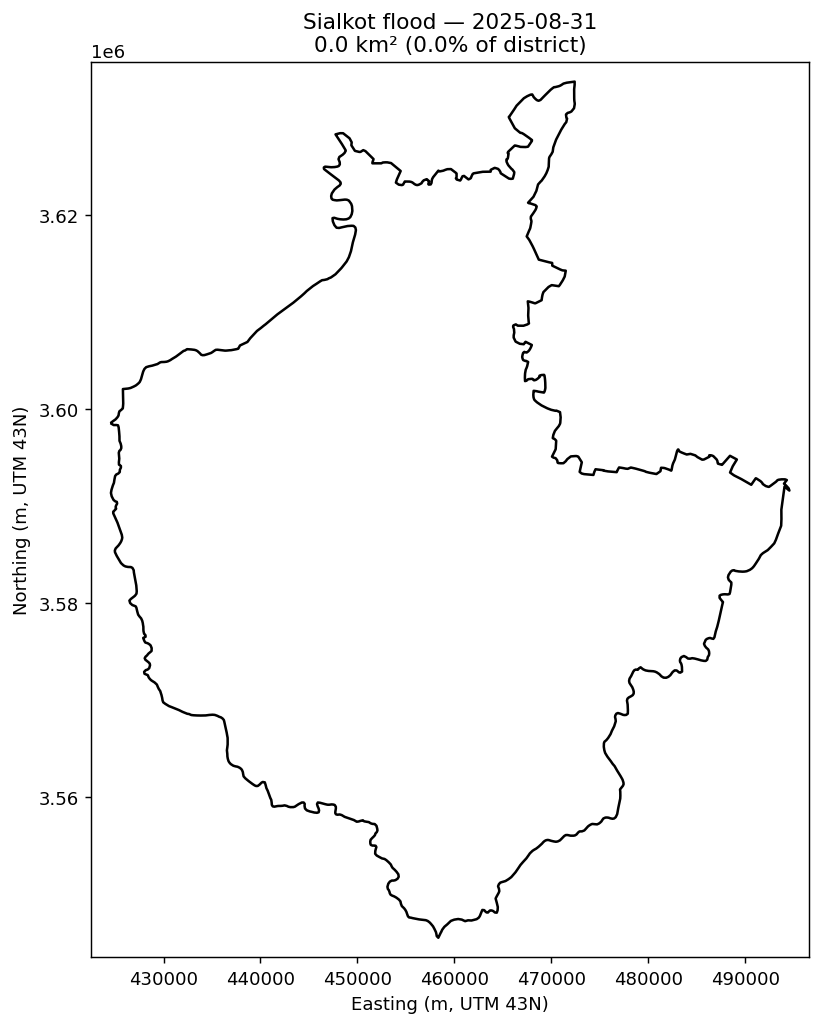}\caption{31 Aug}\end{subfigure}

  \vspace{0.4em}

  \begin{subfigure}[b]{\dwidth}\includegraphics[width=\linewidth]{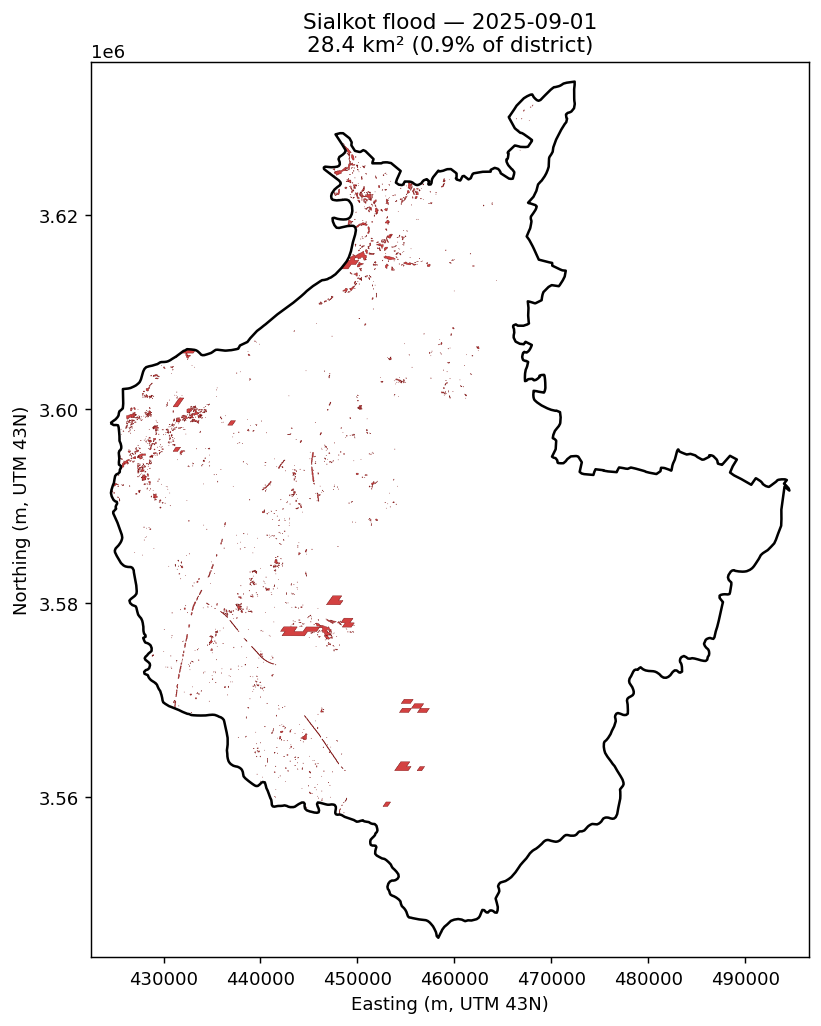}\caption{1 Sep}\end{subfigure}\hfill
  \begin{subfigure}[b]{\dwidth}\includegraphics[width=\linewidth]{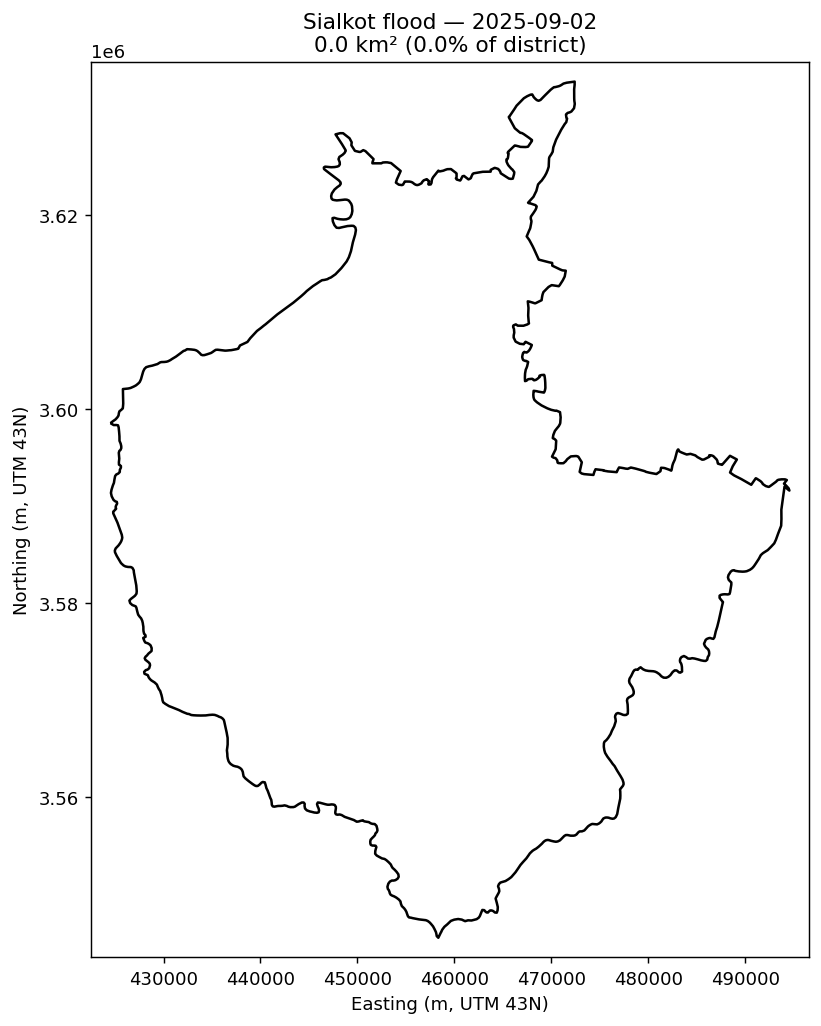}\caption{2 Sep}\end{subfigure}\hfill
  \begin{subfigure}[b]{\dwidth}\includegraphics[width=\linewidth]{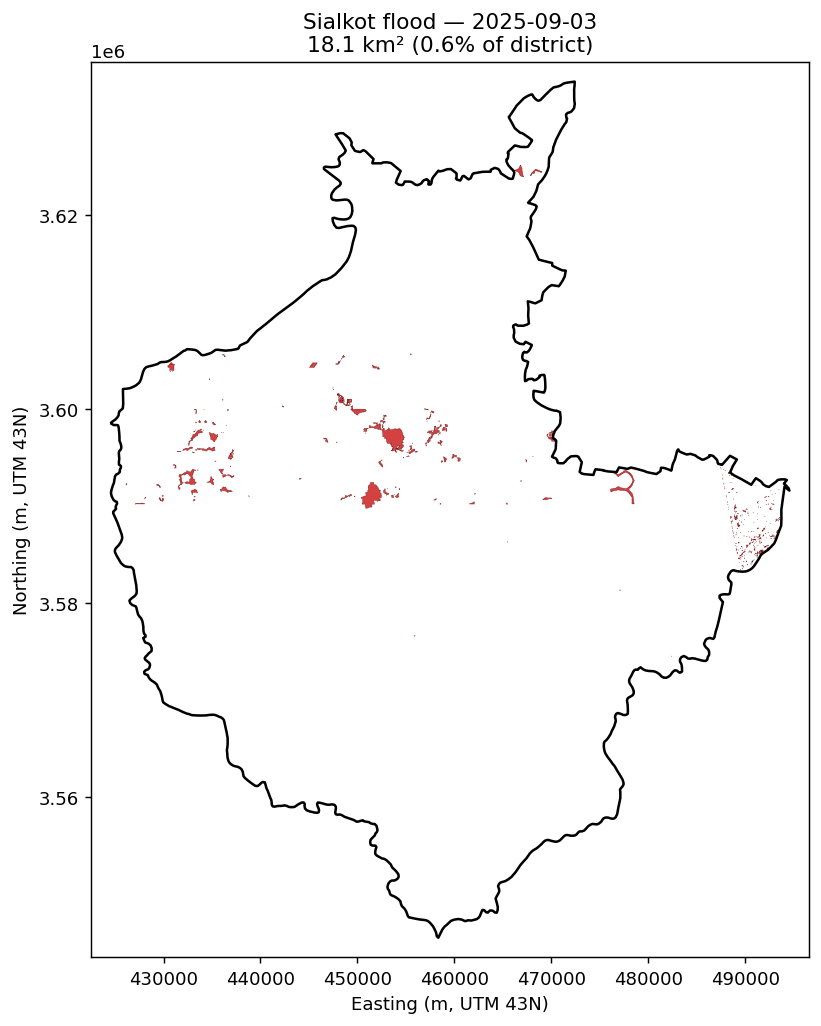}\caption{3 Sep}\end{subfigure}\hfill
  \begin{subfigure}[b]{\dwidth}\includegraphics[width=\linewidth]{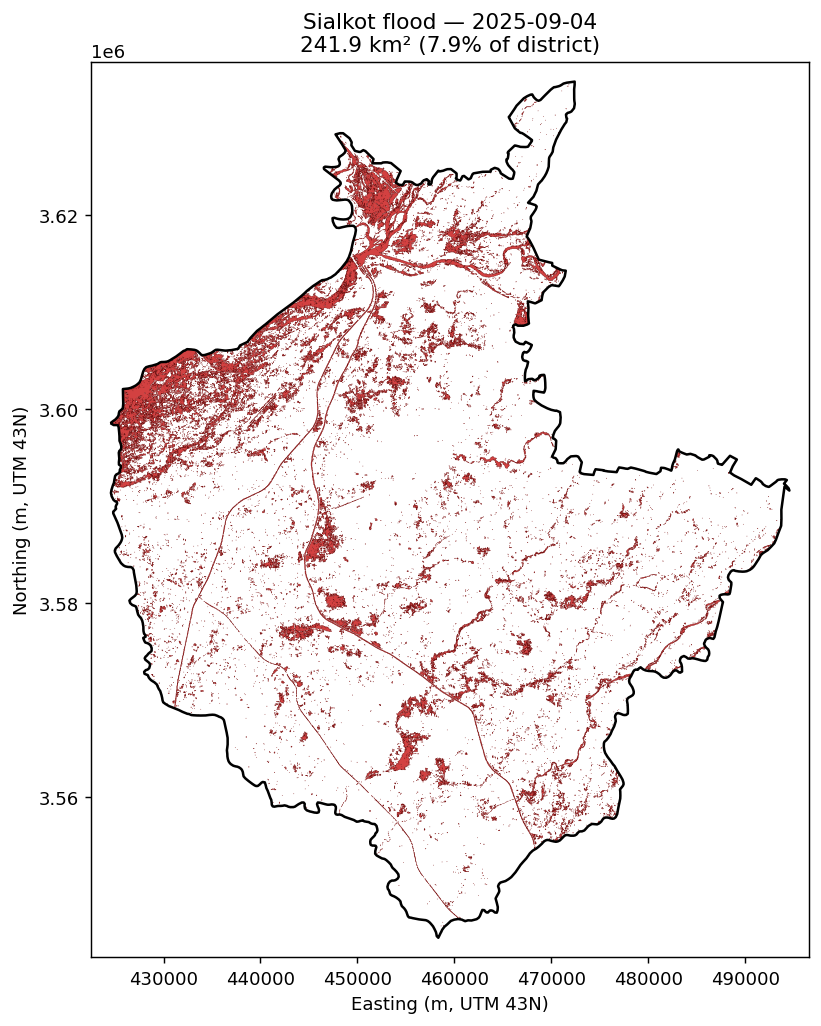}\caption{4 Sep}\end{subfigure}\hfill
  \begin{subfigure}[b]{\dwidth}\includegraphics[width=\linewidth]{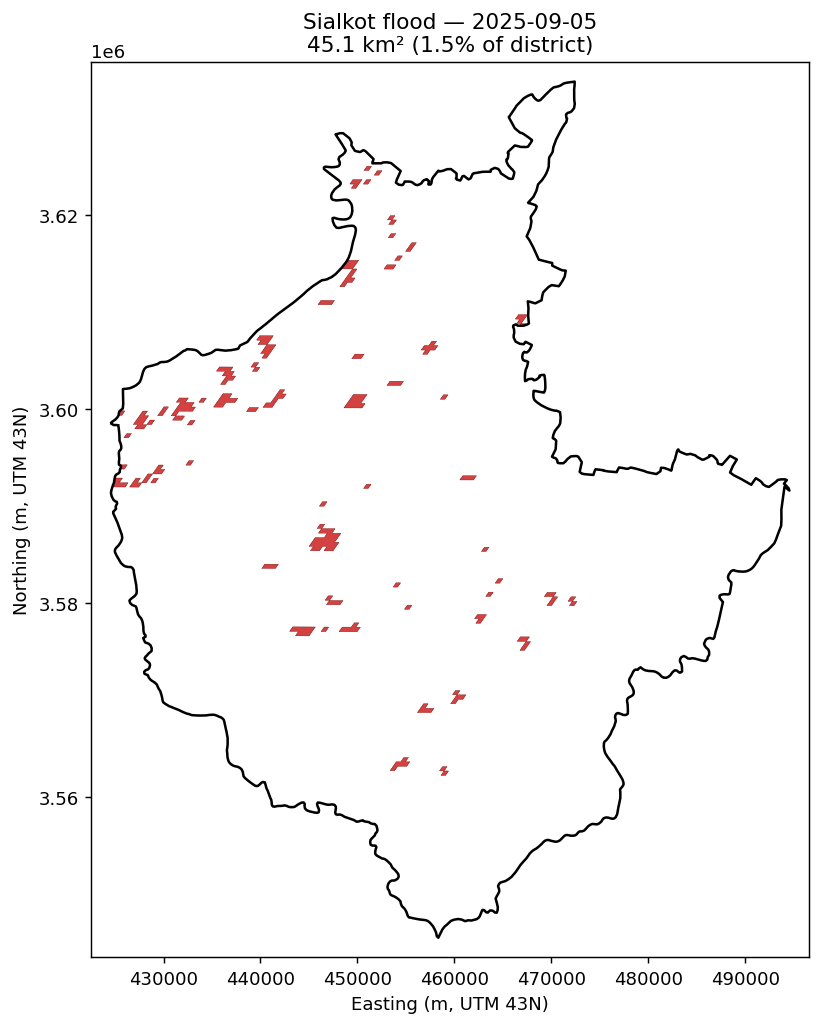}\caption{5 Sep}\end{subfigure}\hfill
  \begin{subfigure}[b]{\dwidth}\includegraphics[width=\linewidth]{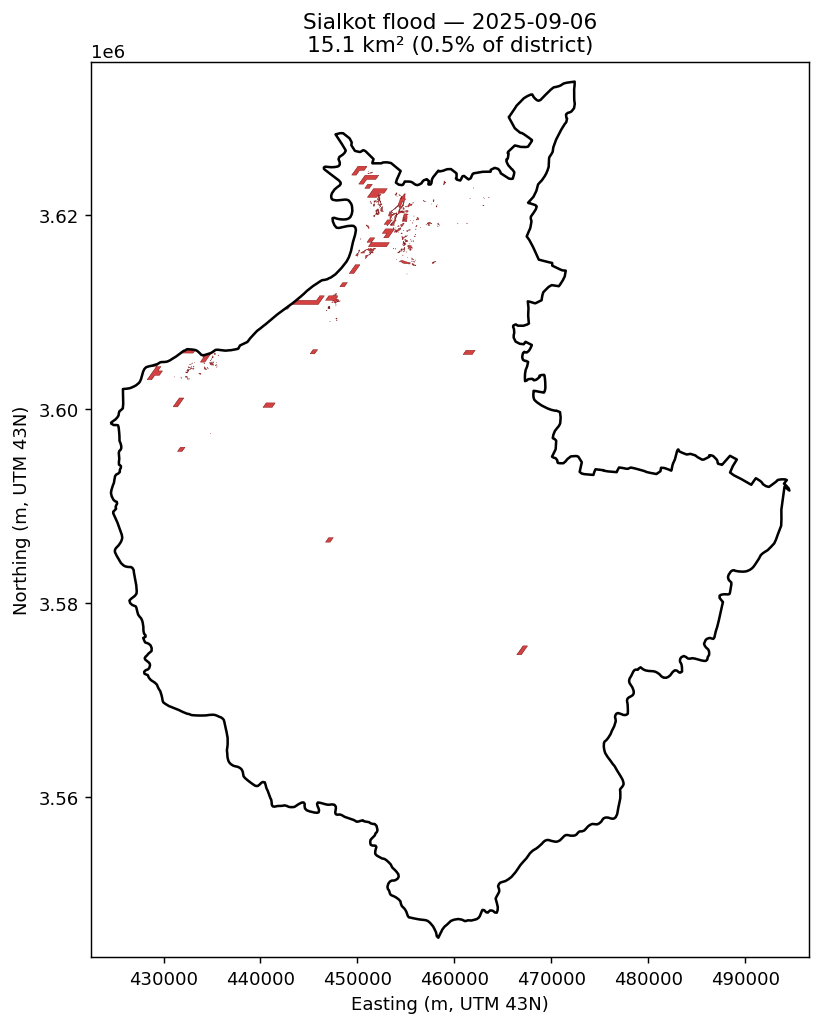}\caption{6 Sep}\end{subfigure}\hfill
  \begin{subfigure}[b]{\dwidth}\includegraphics[width=\linewidth]{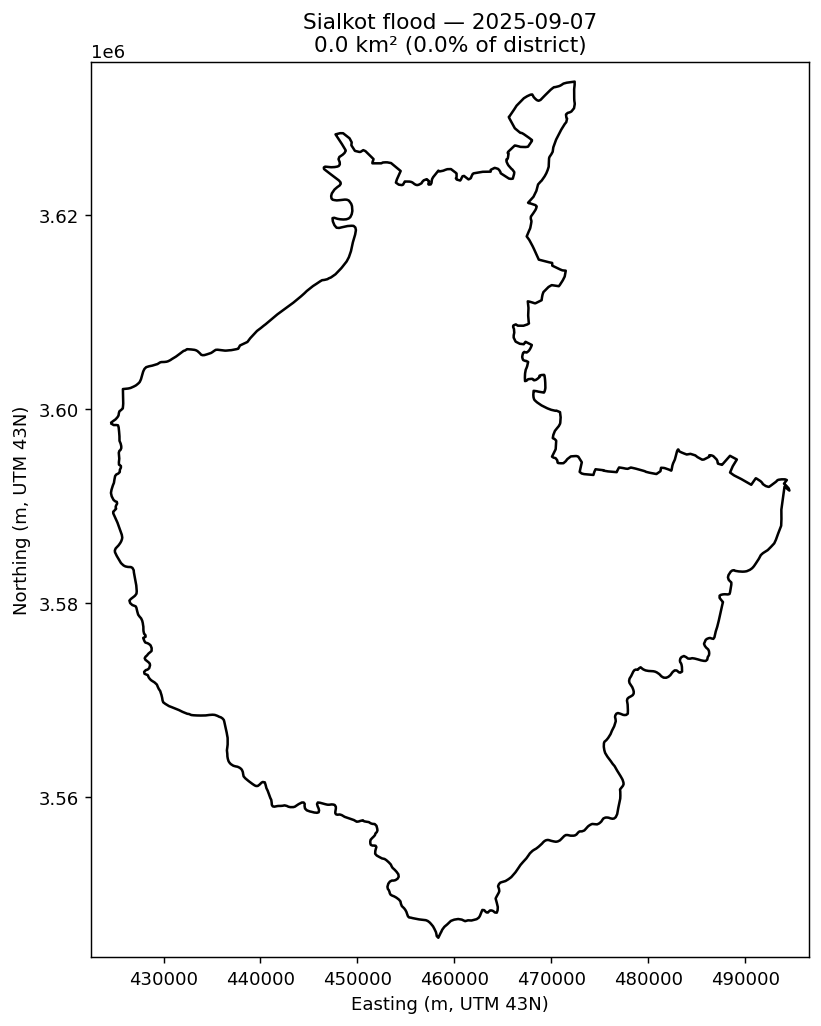}\caption{7 Sep}\end{subfigure}

  \vspace{0.4em}

  \begin{subfigure}[b]{\dwidth}\includegraphics[width=\linewidth]{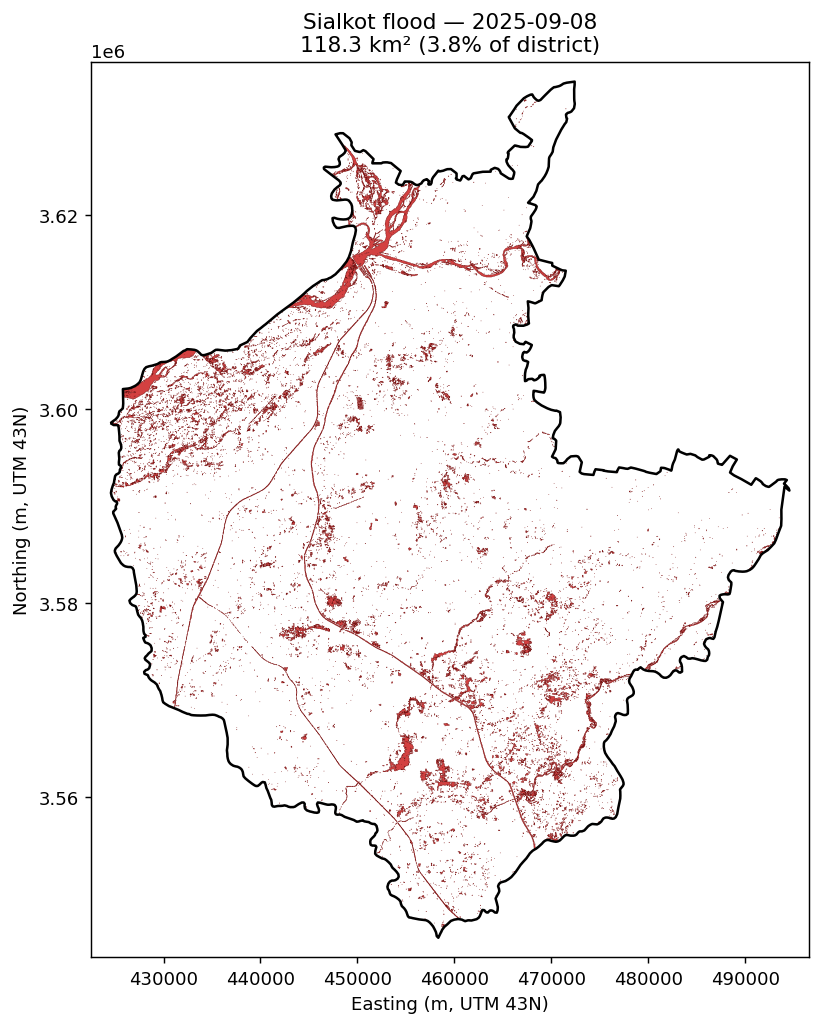}\caption{8 Sep}\end{subfigure}\hfill
  \begin{subfigure}[b]{\dwidth}\includegraphics[width=\linewidth]{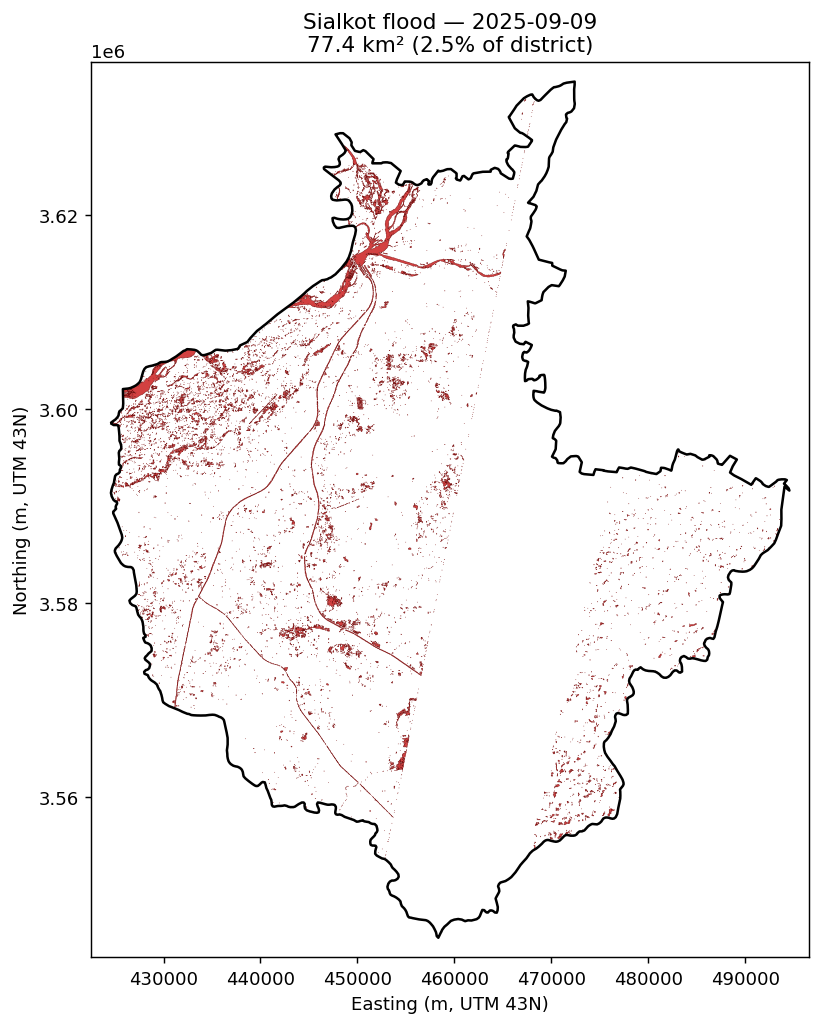}\caption{9 Sep}\end{subfigure}\hfill
  \begin{subfigure}[b]{\dwidth}\includegraphics[width=\linewidth]{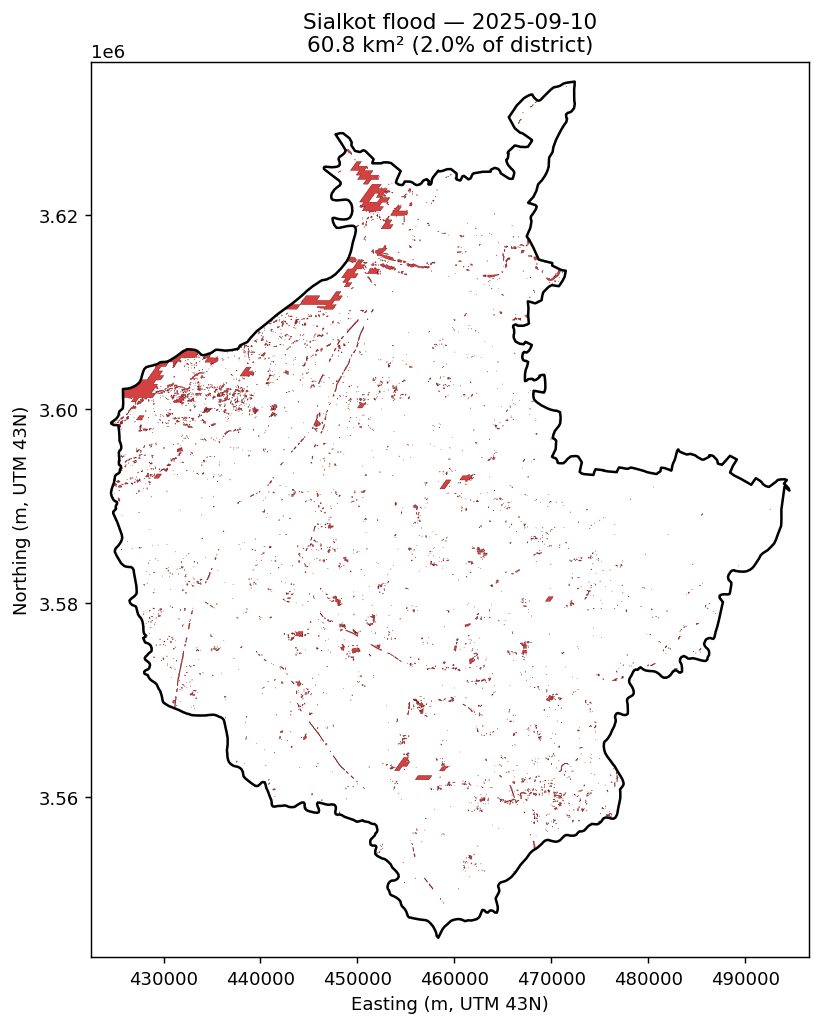}\caption{10 Sep}\end{subfigure}\hfill
  \begin{subfigure}[b]{\dwidth}\includegraphics[width=\linewidth]{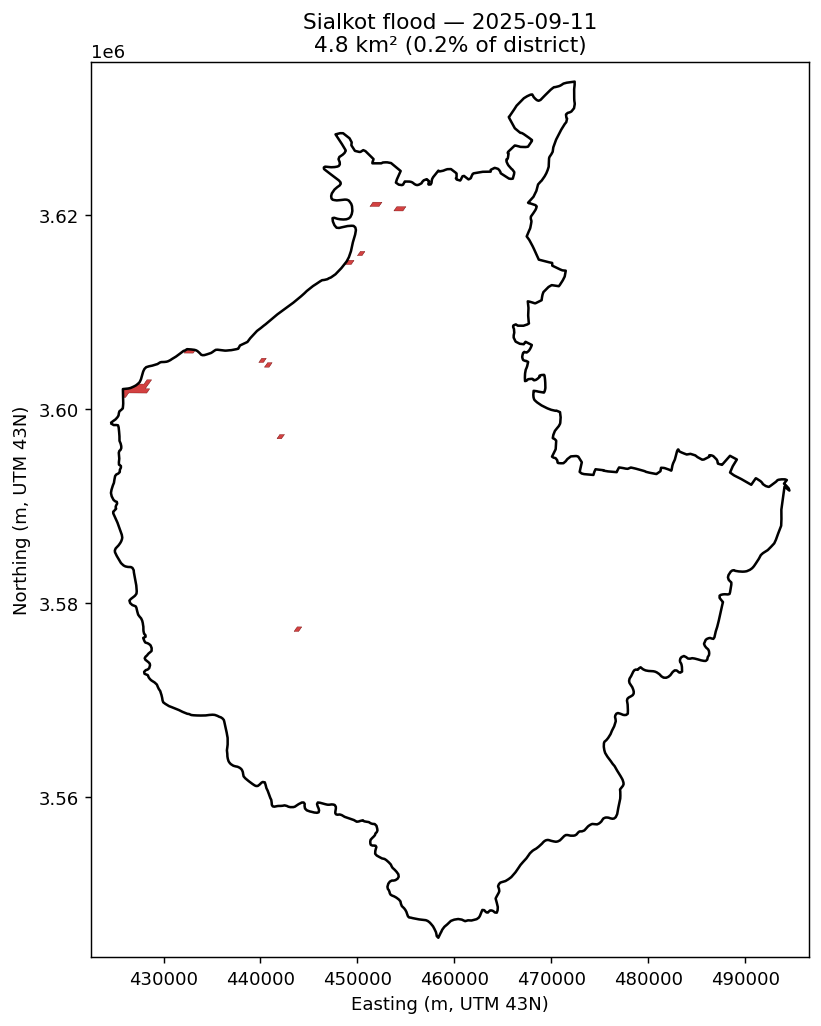}\caption{11 Sep}\end{subfigure}\hfill
  \begin{subfigure}[b]{\dwidth}\includegraphics[width=\linewidth]{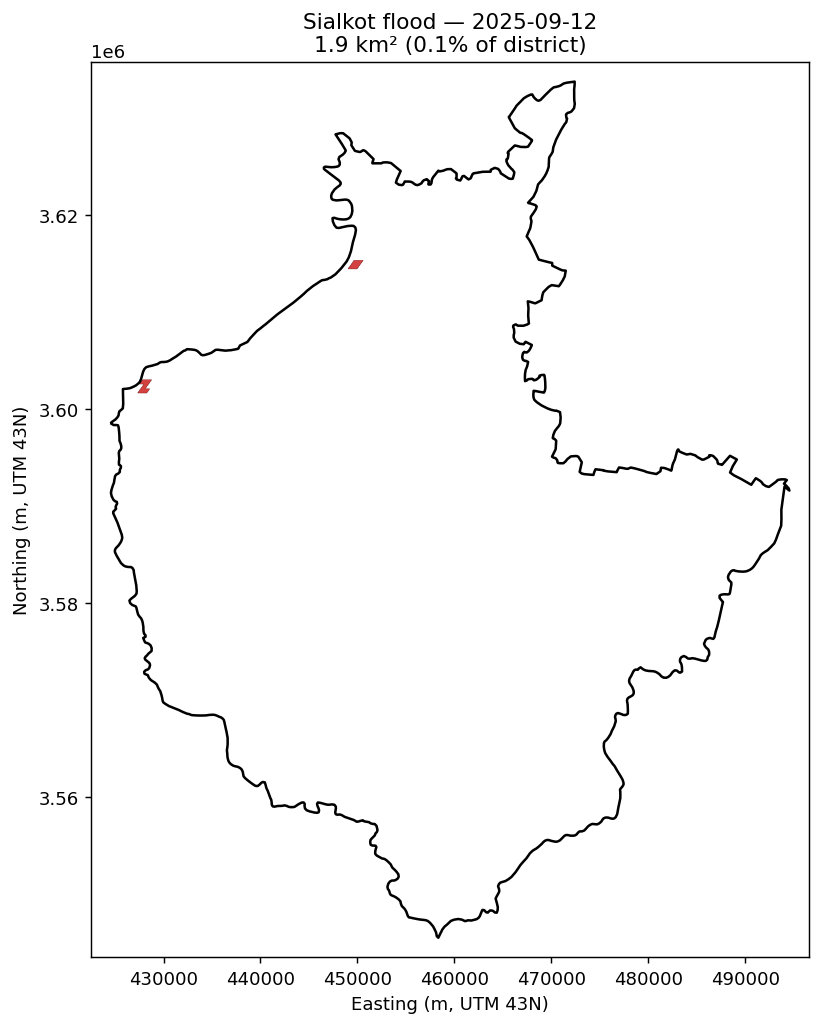}\caption{12 Sep}\end{subfigure}\hfill
  \begin{subfigure}[b]{\dwidth}\includegraphics[width=\linewidth]{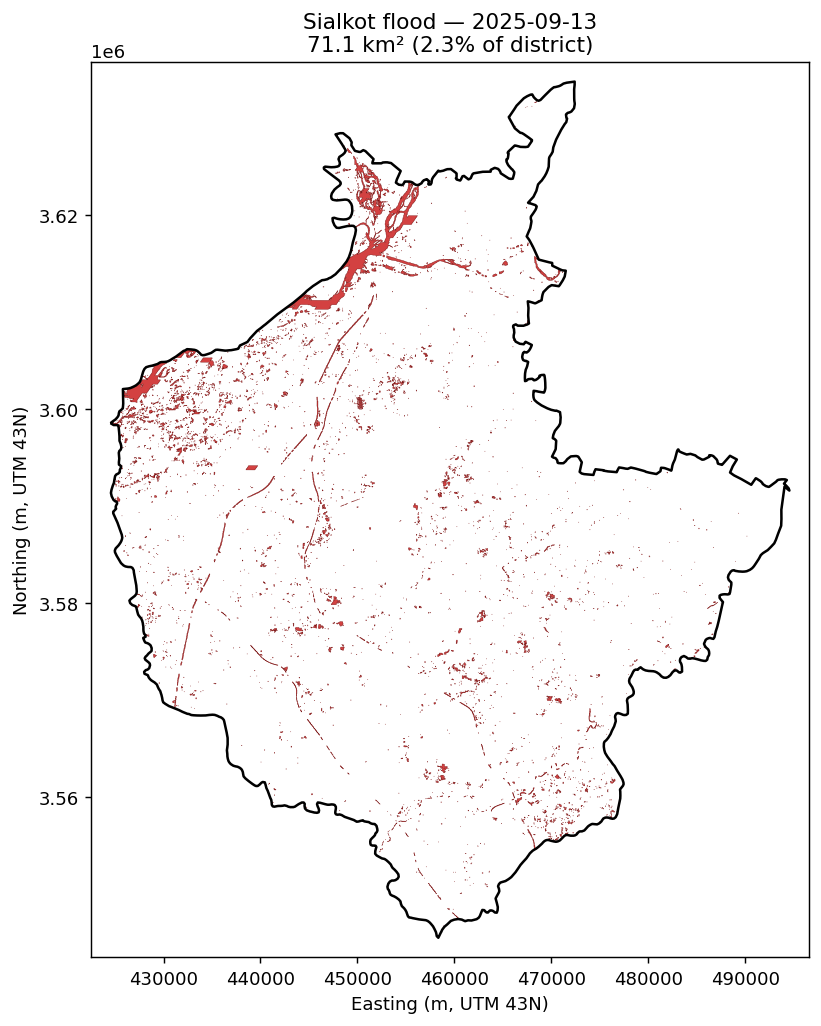}\caption{13 Sep}\end{subfigure}\hfill
  \begin{subfigure}[b]{\dwidth}\includegraphics[width=\linewidth]{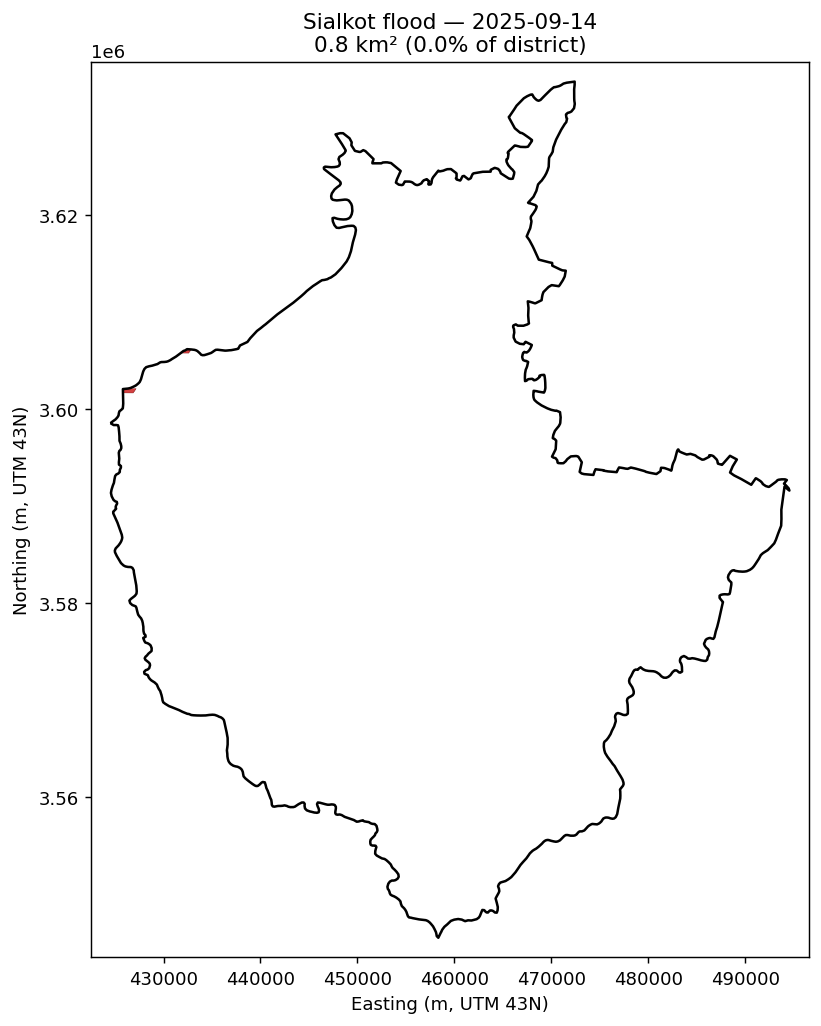}\caption{14 Sep}\end{subfigure}

  \vspace{0.7em}

  \begin{subfigure}[b]{0.42\linewidth}
    \includegraphics[width=\linewidth]{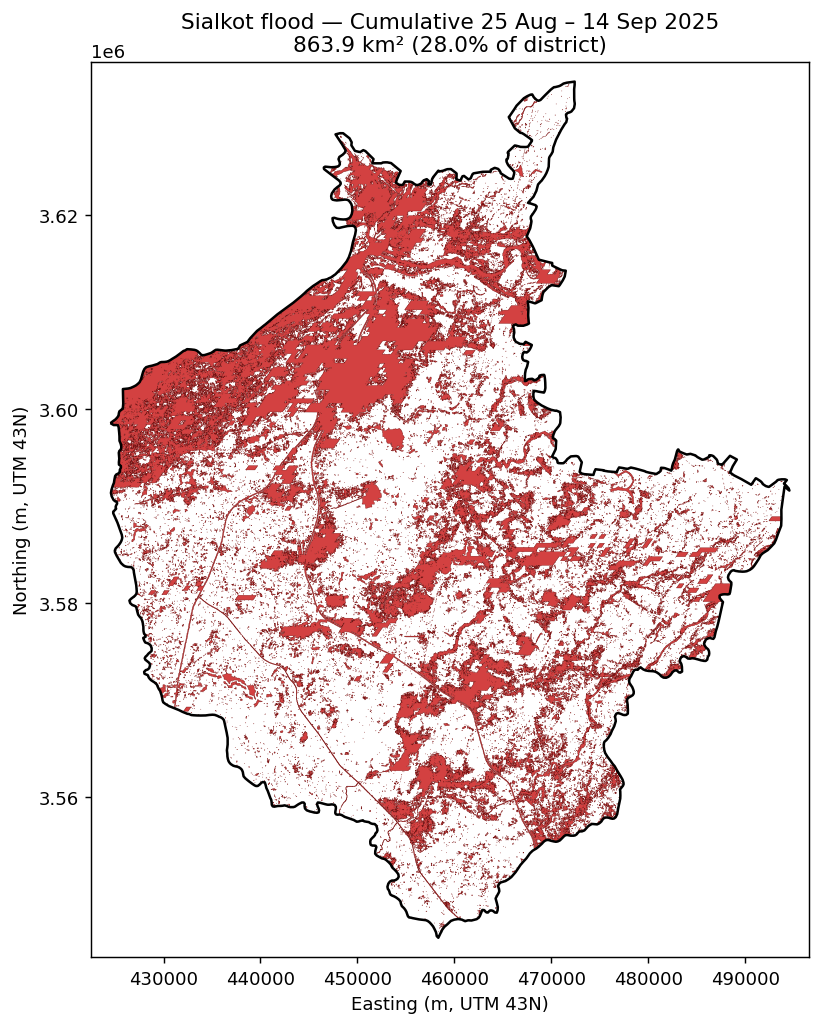}
    \caption{Cumulative flood, 25 Aug -- 14 Sep 2025 (863.9~km$^2$, 28.0\% of district)}
  \end{subfigure}

  \caption{%
    Operational flood nowcasting in Sialkot district through fusion of
    \textbf{five satellite sensors}: HLS Sentinel-2, Sentinel-1 SAR,
    HLS Landsat-30, VIIRS, and MODIS.
    Each daily panel (25~August~--~14~September~2025) shows the inundation
    footprint produced by the logical union of water detections across all
    five sensors that imaged Sialkot on that day. The bottom panel is the
    cumulative flood extent obtained by taking the union of all 21 daily
    masks (863.9~km$^2$; 28.0\% of the district), which is in close
    agreement with the independent cumulative reference product.
    The complementary revisit times and spectral capabilities of the five
    sensors fill in observation gaps caused by clouds and limited single-sensor
    coverage, demonstrating that our multi-sensor fusion approach enables
    \textbf{near-real-time flood nowcasting} at daily granularity throughout
    a multi-week event.%
  }
  \label{fig:sialkot_daily_flood}
\end{figure*}

To illustrate the operational behaviour of the framework, we present a
focused case study on the Pakistan super-flood event that peaked between
Aug 26 - Sept 02, 2025. 
We ran the tiered nowcasting ensemble (Mode A,
Section~\ref{sec:ensemble}) with the rolling pre-/post-flood compositing
windows described in Section~\ref{sec:methods}, stepping the nowcast
timestamp $t$ forward every 48 hours as new Sentinel-1 and HLS
acquisitions became available. At each update, the tiered preference
rule was re-applied pixel-wise over the Pakistan national footprint: on
dates when Tier-1 sensors had observed, the nowcast was driven by
S1/HLS at 30\,m; on dates when only MODIS or VIIRS was available,
Tier-2 observations maintained continuity of the nowcast at their
native resolution. We compare the resulting time series of flood-extent
maps with: (i) the UNOSAT episode footprint published for
26~August--07~September 2025; (ii) GloFAS discharge exceedance along the
Indus and its major tributaries; and (iii) daily precipitation anomalies
from CHIRPS v3.0 \citep{funk2015}.
Figure~\ref{fig:sialkot_daily_flood} summarizes the result.

At the event peak, the tiered nowcast matches the UNOSAT footprint to
within \emph{[insert IoU / overlap metric]}, detecting
16{,}838\,km$^{2}$ of inundation against UNOSAT's 17{,}000\,km$^{2}$
(Table~\ref{tab:unosat}). Beyond the episode window, the retrospective
3/5 majority-vote mosaic (Mode B) over the full JJASO 2025 window
yields a cumulative Pakistan extent of 114{,}513\,km$^{2}$---nearly
seven times the UNOSAT episode footprint---without any additional manual
activation. Together, the two modes demonstrate that the framework can
serve both as a near-real-time nowcast during a designated emergency
(Mode A) and as a continuous seasonal monitor between episodes (Mode B).

\subsection{Nowcasting Case Study 2: Khyber Pakhtunkhwa (KP)}
\label{sec:nowcast_case2}

For the Khyber Pakhtunkhwa (KP) case study, we generated a coupled multi-hazard product spanning the peak of the 2025 monsoon season, covering 1~August through 19~September~2025. The product jointly characterises three of the dominant hydro-geomorphic processes affecting the province---flash flooding, glacial lake outburst floods (GLOFs), and rainfall-triggered landslides---which frequently occur in cascade across KP's steep northern catchments and were responsible for the majority of reported losses during the study window. Composite hazard maps were produced at five-day intervals (ten time slices) together with a cumulative product integrating the full 46-day window, allowing both the temporal evolution and the aggregate spatial footprint of compound hazards to be examined. Daily flash and landslide flooding footprints are additionally presented in appendix~\ref{casestudy2kp}.

\FloatBarrier

\section{Limitations and Broader Impact}
\label{sec:limitations}

\paragraph{Limitations.}
Our framework inherits the intrinsic limits of its inputs. Optical sensors
(HLS L30/S30, MODIS, VIIRS) can be degraded by residual cloud and cirrus
contamination during monsoon peaks, reducing the effective revisit rate
for per-pixel change detection. Sentinel-1 SAR is sensitive to flooded
vegetation, urban double-bounce backscatter, and wet bare soils, which can
cause both omissions and commissions in densely built or forested terrain.
The tiered nowcasting ensemble (Mode A) preserves high resolution when
Tier-1 sensors are available, but falls back to 500\,m MODIS or
375--750\,m VIIRS when they are not; nowcast pixels driven by these
fallback sensors therefore carry genuinely coarser spatial detail than
Tier-1--driven pixels, and downstream exposure calculations at those
pixels inherit that coarseness. We output an observation-mask layer
recording which tier produced each nowcast pixel so that users can
restrict analyses to Tier-1--driven areas when spatial precision matters
more than timeliness. For the retrospective seasonal product (Mode B),
the 3-of-5 majority-voting rule trades a small number of false negatives
for higher agreement; a tuned probabilistic aggregation would likely
improve recall in data-sparse regions. UNOSAT ground-truth
coverage is available for only three countries in our region and for
limited time windows, so our quantitative validation against external
products is episode-level rather than continuous. Population exposure
uses gridded estimates (WorldPop 2025) that are themselves modelled
products and inherit their own uncertainty, which we do not propagate
into the reported counts.

\paragraph{Broader impact.}
Continuous, harmonized flood-extent products can directly support
humanitarian response, damage assessment, and climate-adaptation planning
across South Asia---a region where over a quarter of a billion people were
exposed to flooding in 2025 alone. Open, reproducible pipelines reduce the
reliance of national agencies on ad hoc episode-triggered mapping and
shrink the gap between high-capacity and resource-limited jurisdictions.

At the same time, over-reliance on remotely sensed flood extents carries
risks. False negatives in urban or vegetated areas can leave exposed
populations uncounted in damage assessments; false positives can misdirect
resources. Our maps should be used alongside, not in place of, on-the-ground
assessments, and we explicitly caution against using them for
settlement-level allocation decisions without local validation. The
population-exposure figures reported here are indicative regional
estimates and should not be interpreted as individual-level or
household-level inferences.


\FloatBarrier
\section{Discussion and Conclusion}

We have introduced a multi-sensor ensemble remote-sensing framework that
delivers spatially and temporally continuous flood extent mapping over an
entire monsoon season in Pakistan. Strong agreement with UNOSAT episode-level products supports the reliability of the framework at the event scale. The continuous view recovers a substantial additional flood extent that is missed by episode-triggered products, with direct implications for aggregated exposure estimates. Limitations include residual optical-sensor cloud contamination, SAR sensitivity to urban surfaces and vegetation, and remaining coverage gaps for countries where UNOSAT products are unavailable. Future work will explore probabilistic ensemble weighting, uncertainty quantification, and integration with near-real-time operational products for humanitarian response.

\section*{Data Availability}

Built-up area and cropland estimates are derived from ESA WorldCover 2021
(\url{https://developers.google.com/earth-engine/datasets/catalog/ESA_WorldCover_v200}).
Road, school, and hospital layers are from Giga-HOTOSM
(\url{https://www.hotosm.org/en/}).



\bibliographystyle{plainnat}
\bibliography{references}

\appendix

\section{Key Terms and Definitions}
\label{app:defs}

\begin{description}
  \item[Hazard.] The physical event itself, described in terms of
    probability and intensity.
  \item[Exposure.] The people, infrastructure, and ecosystems located in
    harm's way.
  \item[Vulnerability.] The susceptibility of exposed people and systems
    to harm from a physical event.
  \item[Mitigation.] Efforts to reduce greenhouse gas emissions or enhance
    the sinks that absorb them.
  \item[Adaptation.] Efforts to reduce vulnerability through better
    infrastructure, warning systems, and social protections, or to reduce
    exposure through managed retreat.
  \item[Climatic Impact Driver.] A physical climate condition that
    directly affects society or ecosystems.
\end{description}

\section{Casestudy 2: Khyber Pakhtunkhwa (KP)}\label{casestudy2kp}

\begin{figure}[H]
    \centering
    \foreach \d [count=\i] in {01,02,03,04,05,06,07,08,09,10,%
                               11,12,13,14,15}{%
        \begin{subfigure}[t]{0.23\textwidth}
            \centering
            \includegraphics[width=\linewidth]{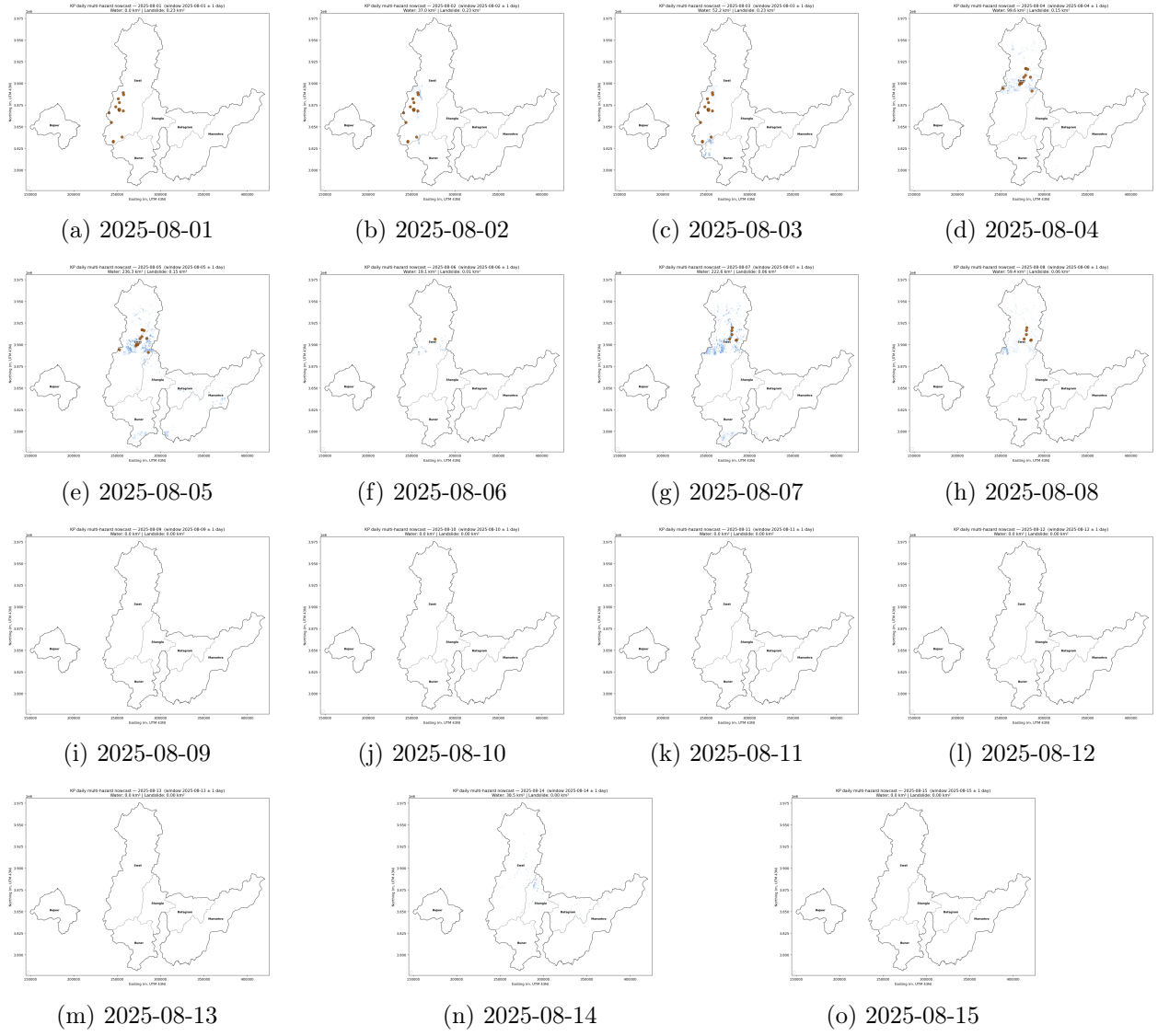}
            \caption{2025-08-\d}
            \label{fig:kp_hazard_2025-08-\d}
        \end{subfigure}%
        \ifnum\numexpr\i\relax=\numexpr(\i/4)*4\relax\par\medskip\else\hfill\fi
    }
    \caption{Multi-hazard maps for Khyber Pakhtunkhwa, August 2025 (days 1–15).}
    \label{fig:kp_hazard_aug2025_a}
\end{figure}

\begin{figure}[H]
    \centering
    \foreach \d [count=\i] in {16,17,18,19,20,%
                               21,22,23,24,25,26,27,28,29,30,31}{%
        \begin{subfigure}[t]{0.23\textwidth}
            \centering
            \includegraphics[width=\linewidth]{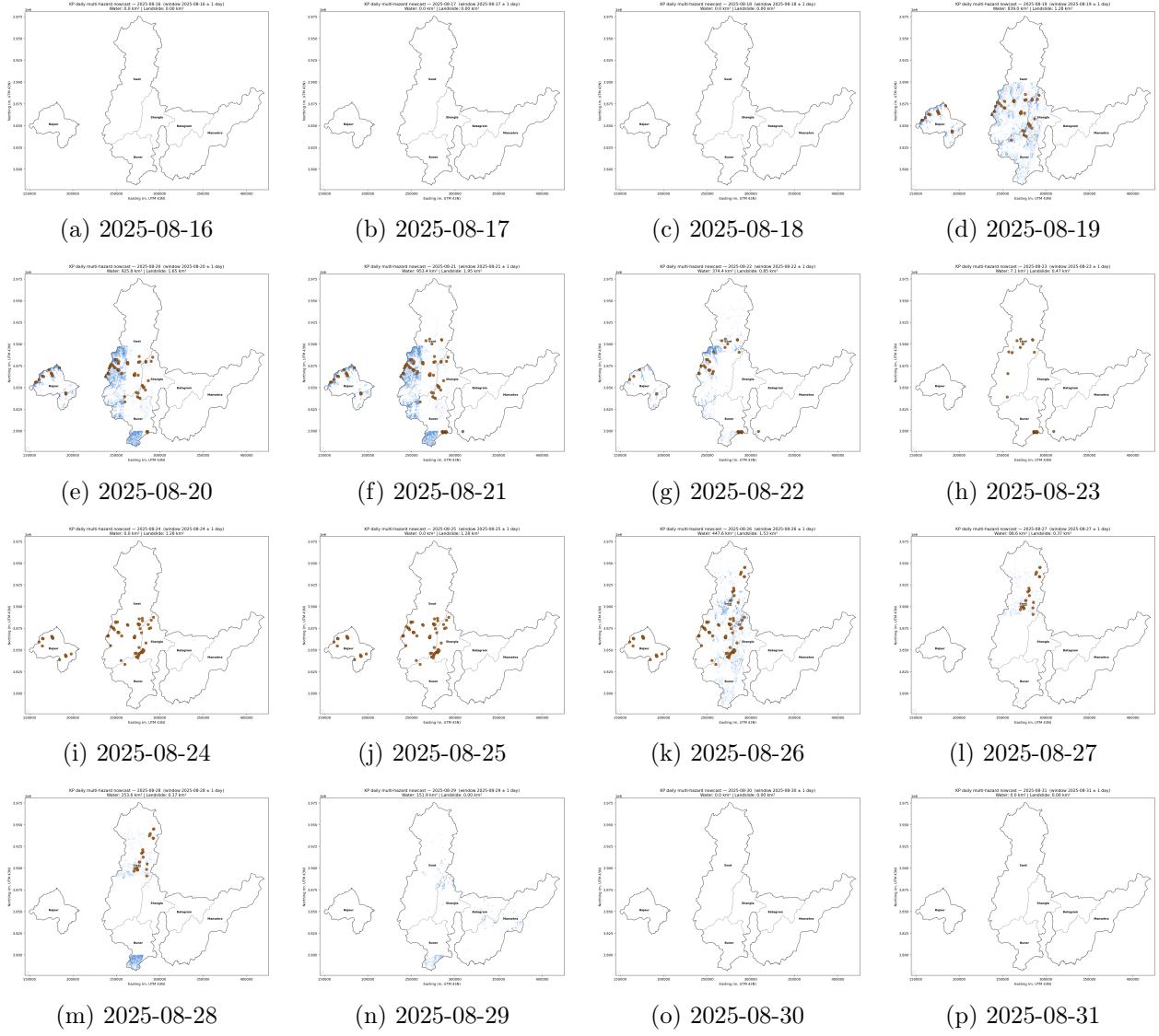}
            \caption{2025-08-\d}
            \label{fig:kp_hazard_2025-08-\d}
        \end{subfigure}%
        \ifnum\numexpr\i\relax=\numexpr(\i/4)*4\relax\par\medskip\else\hfill\fi
    }
    \caption{Multi-hazard maps for Khyber Pakhtunkhwa, August 2025 (days 16–31).}
    \label{fig:kp_hazard_aug2025_b}
\end{figure}

\begin{figure}[H]
    \centering
    \foreach \d [count=\i] in {01,02,03,04,05,06,07,08,09,10,%
                               11,12,13,14,15}{%
        \begin{subfigure}[t]{0.23\textwidth}
            \centering
            \includegraphics[width=\linewidth]{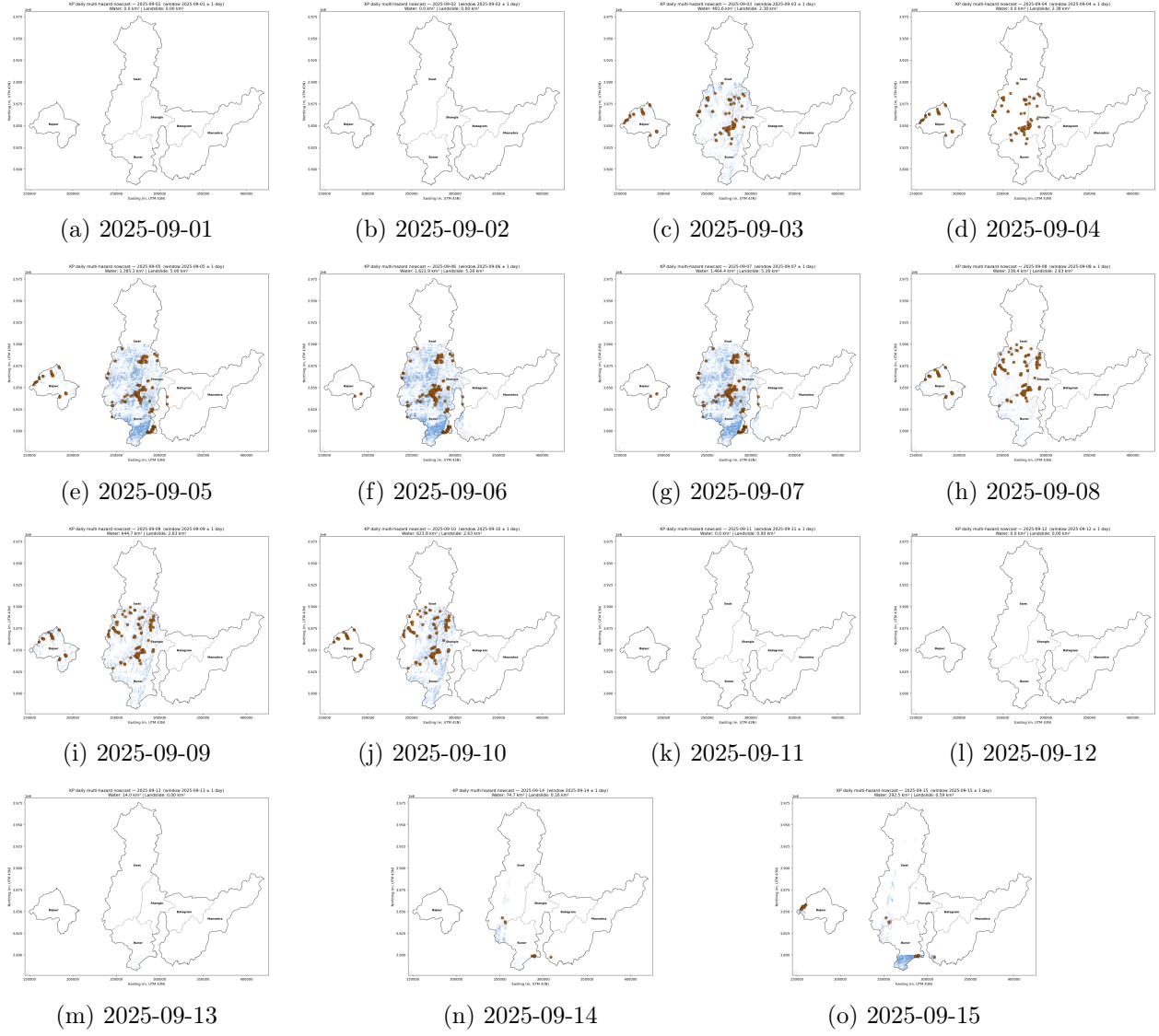}
            \caption{2025-09-\d}
            \label{fig:kp_hazard_2025-09-\d}
        \end{subfigure}%
        \ifnum\numexpr\i\relax=\numexpr(\i/4)*4\relax\par\medskip\else\hfill\fi
    }
    \caption{Multi-hazard maps for Khyber Pakhtunkhwa, September 2025 (days 1–15).}
    \label{fig:kp_hazard_sep2025_a}
\end{figure}

\begin{figure}[H]
    \centering
    \foreach \d [count=\i] in {16,17,18,19,20,%
                               21,22,23,24,25,26,27,28,29,30}{%
        \begin{subfigure}[t]{0.23\textwidth}
            \centering
            \includegraphics[width=\linewidth]{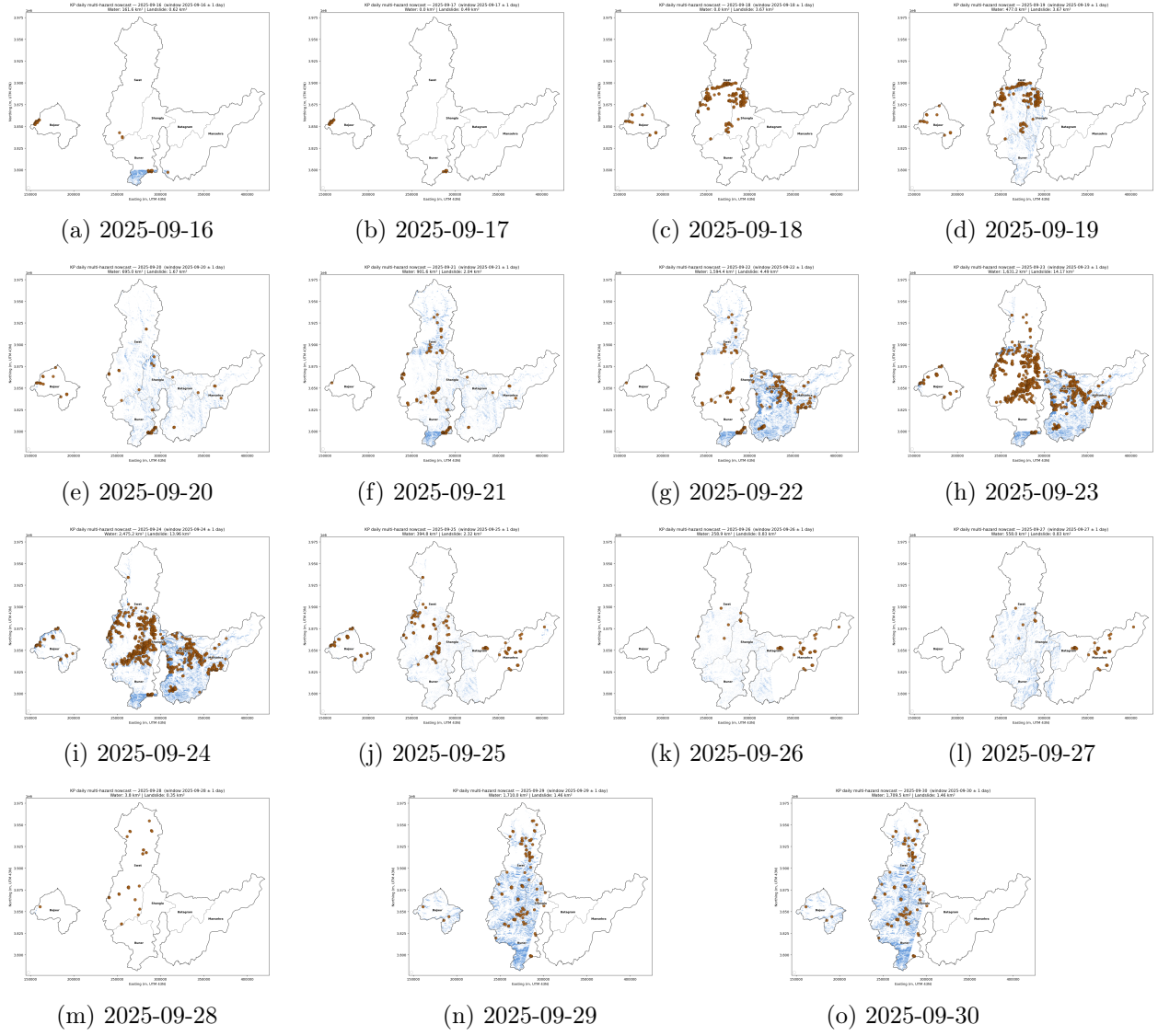}
            \caption{2025-09-\d}
            \label{fig:kp_hazard_2025-09-\d}
        \end{subfigure}%
        \ifnum\numexpr\i\relax=\numexpr(\i/4)*4\relax\par\medskip\else\hfill\fi
    }
    \caption{Multi-hazard maps for Khyber Pakhtunkhwa, September 2025 (days 16–30).}
    \label{fig:kp_hazard_sep2025_b}
\end{figure}

\begin{figure}[H]
    \centering
    \includegraphics[width=0.95\textwidth]{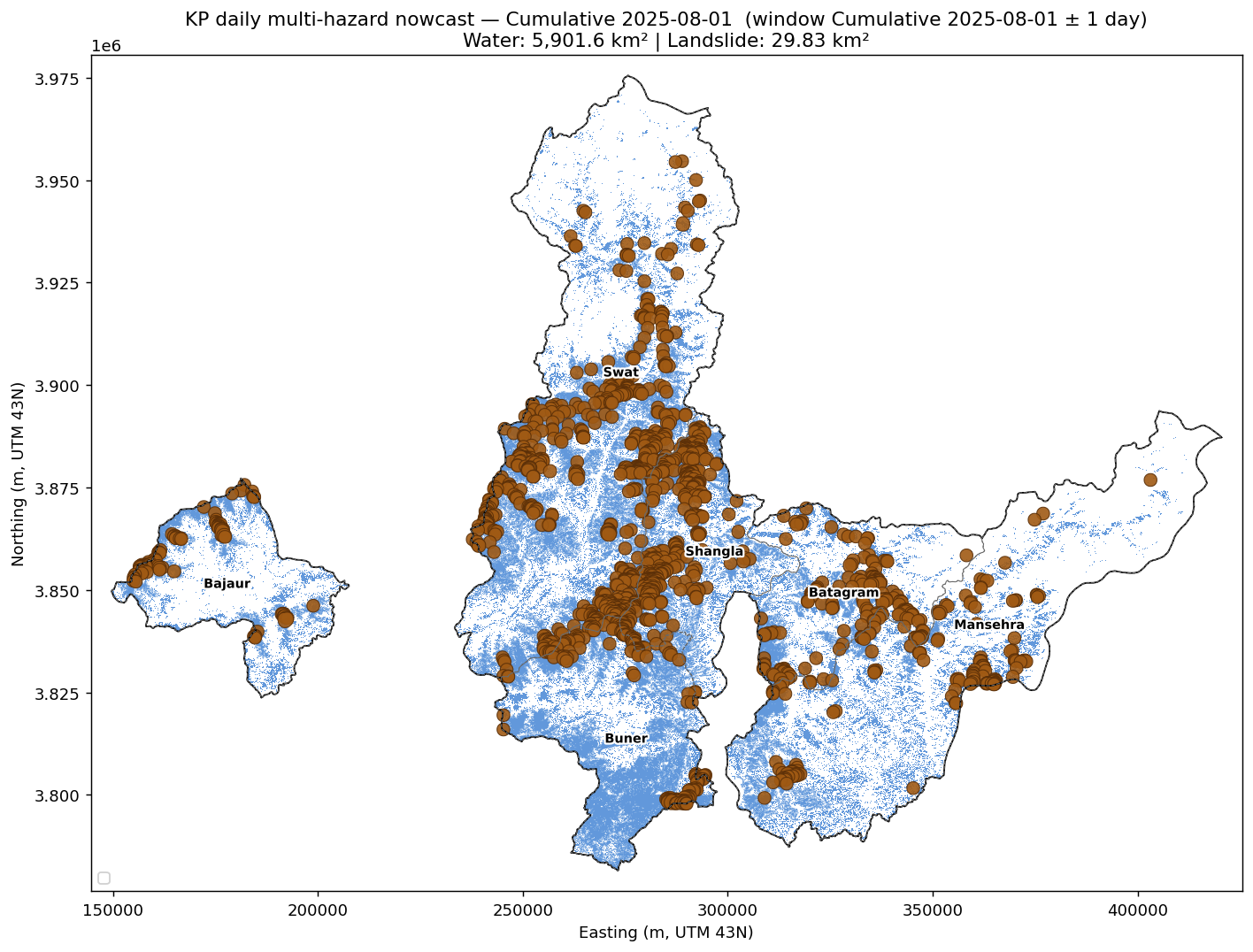}
    \caption[Cumulative multi-hazard footprint, 1~August--30~September~2025]{%
        Cumulative multi-hazard footprint over Khyber Pakhtunkhwa for the
        2025 monsoon study window (1~August--30~September~2025), obtained
        by aggregating the 60 daily hazard composites shown in the
        preceding figures. Each pixel represents the maximum hazard
        intensity recorded across the 60-day window, jointly integrating
        flash-flood, glacial lake outburst flood (GLOF), and
        rainfall-triggered landslide hazards. The map therefore
        delineates the spatial envelope of compound hydro-geomorphic
        hazard exposure during the peak monsoon period and provides the
        basis for the regional impact assessment discussed in the main
        text.}
    \label{fig:kp_hazard_cumulative}
\end{figure}

\end{document}